\documentclass[conference,a4paper,10pt,times]{IEEEtran}

\newif\iffullpaper
\fullpapertrue

\newif\ifshowcomment

\usepackage{array}
\usepackage{fancyhdr}
\usepackage{booktabs}
\usepackage{amsmath}
\usepackage{amsthm}
\usepackage{amsfonts}
\usepackage{amssymb}
\usepackage{calrsfs}
\usepackage{pifont}
\usepackage{verbatim}
\usepackage{url}
\usepackage{hyperref}
\hypersetup{
    colorlinks=true,
    allcolors=blue,
}
\usepackage{graphicx}
\usepackage{sidecap}
\usepackage{float}
\usepackage{enumitem}
\usepackage{color}
\usepackage{listings}
\usepackage{algorithm}
\usepackage{algpseudocode}
\usepackage{multirow}
\usepackage{xspace}
\usepackage[usenames,dvipsnames]{xcolor}

\lstdefinelanguage{rust}{
  keywords={String, u64, HashMap, Address},
  keywordstyle=\color{blue}\bfseries,
  ndkeywords={pub, struct, impl, return, fn, if, for, let},
  ndkeywordstyle=\color{black}\bfseries,
  identifierstyle=\color{black},
  sensitive=false,
  comment=[l]{//},
  morecomment=[s]{/*}{*/},
  commentstyle=\color{purple}\ttfamily,
  stringstyle=\color{red}\ttfamily,
  morestring=[b]',
  morestring=[b]"
}
\lstdefinelanguage{javascript}{
  keywords={typeof, new, true, false, catch, function, return, null, catch, switch, var, if, in, while, do, else, case, break},
  keywordstyle=\color{blue}\bfseries,
  ndkeywords={class, export, boolean, throw, implements, import, this},
  ndkeywordstyle=\color{black}\bfseries,
  identifierstyle=\color{black},
  sensitive=false,
  comment=[l]{//},
  morecomment=[s]{/*}{*/},
  commentstyle=\color{purple}\ttfamily,
  stringstyle=\color{red}\ttfamily,
  morestring=[b]',
  morestring=[b]"
}

\newcommand{\squishlist}{\begin{itemize}[itemsep=0pt,parsep=0pt,topsep=0pt,partopsep=0pt,leftmargin=1em,labelwidth=1em,labelsep=0.5em]}
\newcommand{\squishlistend}{\end{itemize}}
\newcommand{\squishend}{\end{itemize}}
\newcommand{\squishenum}{\begin{enumerate}[itemsep=0.5pt,parsep=0pt,topsep=0pt,partopsep=0pt,leftmargin=1.5em,labelwidth=1em,labelsep=0.5em]{}}
\newcommand{\squishenumend}{\end{enumerate}}
\newcommand{\squishdesc}{\begin{description}[itemsep=0.5pt,parsep=0pt,topsep=0pt,partopsep=0pt,leftmargin=1.5em,labelwidth=1em,labelsep=0.5em]{}}
\newcommand{\squishdescend}{\end{description}}

\newtheorem{theorem}{Theorem}

\usepackage[ff,sets,adversary,keys,primitives,operators,lambda,logic,notions]{cryptocode}
\usepackage[capitalize]{cleveref}
\usepackage{boxedminipage}
\usepackage[utf8x]{inputenc}
\usepackage{float}
\usepackage{nicefrac}
\usepackage{enumitem}
\usepackage{framed}
\usepackage{makecell}

\DeclareMathAlphabet{\mathcal}{OMS}{cmsy}{m}{n}

\newcommand{\onlyinfullversion}[1]{\iffullpaper
\cref{#1}\else
the online version~\cite{ekidenFullVersion}\fi}

\ifshowcomment
\newcommand{\todo}[1]{\textsf{\color{red}{[{TODO: #1}]}}}
\newcommand{\ray}[1]{\textsf{\color{orange}{[Ray: {#1}]}}}
\newcommand{\fanz}[1]{\textsf{\color{blue}{[Fan: {#1}]}}}
\newcommand{\dawn}[1]{\textsf{\color{red}{[Dawn: {#1}]}}}
\newcommand{\ari}[1]{\textsf{\color{blue}{[Ari: {#1}]}}}
\newcommand{\anote}[1]{\textsf{\color{red}{[Andrew: {#1}]}}}
\newcommand{\wh}[1]{\textsf{\color{red}{[Warren: {#1}]}}}
\else
\newcommand{\todo}[1]{}
\newcommand{\ray}[1]{}
\newcommand{\fanz}[1]{}
\newcommand{\dawn}[1]{}
\newcommand{\ari}[1]{}
\newcommand{\anote}[1]{}
\newcommand{\wh}[1]{}
\fi

\colorlet{party}{Brown}
\colorlet{protocol}{Black}
\colorlet{string}{BlueViolet}
\definecolor{DarkGreen}{rgb}{0.0, 0.4, 0.0}
\colorlet{entry}{NavyBlue}

\def\systemnameRaw{Ekiden}
\def\systemname{\systemnameRaw\xspace}
\def\systemimpl{\systemnameRaw-BT\xspace}

\newcommand{\smartcontract}{confidentiality-preserving smart contract\xspace}
\newcommand{\smartcontracts}{confidentiality-preserving smart contracts\xspace}

\newcommand{\ekdparty}{\textcolor{party}{\mathcal{P}}}
\newcommand{\ekdpartyi}{\textcolor{party}{\mathcal{P}_i}}
\newcommand{\ekdpartysig}{\ensuremath{\sigma_{\ekdpartyi}}}
\newcommand{\ekdadv}{\textcolor{party}{\adv}}
\newcommand{\ekdprog}{{\mathsf{Contract}}}

\newcommand{\ekdstate}{\ensuremath{\mathsf{st}}}
\newcommand{\ekdstateprev}{\ensuremath{\ekdstate_\text{old}}}
\newcommand{\ekdstateprevct}{\ensuremath{\enc(\seckeycont,\ekdstateprev)}}
\newcommand{\ekdstatenew}{\ensuremath{\ekdstate_\text{new}}}
\newcommand{\ekdstatenewct}{\ensuremath{\enc(\seckeycont,\ekdstatenew)}}
\newcommand{\ekdstateprevhash}{\ensuremath{h_\text{old}}}
\newcommand{\ekdstatect}{\ensuremath{\ekdstate_{\text{ct}}}}
\newcommand{\ekdstatediff}{\ensuremath{\Delta\ekdstate}}
\newcommand{\ekdstatediffct}{\ensuremath{\ekdstatediff_{\text{ct}}}}

\newcommand{\ekdatt}{\ensuremath{\sigma_{\text{\tee}}}}
\newcommand{\sigatt}{\ensuremath{\pi}}
\newcommand{\sigias}{\ensuremath{\sigma_{\text{IAS}}}}

\newcommand{\ekdstorage}{{\mathsf{Storage}}}

\newcommand{\ekdinput}{{\mathsf{inp}}}
\newcommand{\ekdinputct}{{\ensuremath{\mathsf{inp}_\text{ct}}}}
\newcommand{\ekdinputhash}{\ensuremath{h_\ekdinput}}
\newcommand{\ekdoutput}{{\mathsf{outp}}}
\newcommand{\ekdoutputct}{\ensuremath{\mathsf{outp}_\text{ct}}}
\newcommand{\ekdoutputhash}{\ensuremath{h_\ekdoutput}}
\newcommand{\ekdid}{\mathsf{id}}
\newcommand{\ekdvalue}{\mathsf{val}}
\newcommand{\ekdcid}{\textcolor{entry}{\mathsf{cid}}}
\newcommand{\enclaveid}{\textcolor{entry}{\mathsf{eid}}}

\newcommand{\ekdenv}{\ensuremath{\textcolor{party}{\mathcal{Z}}}}
\newcommand{\ekdstatetrans}{\ensuremath{\mathsf{trans}}}
\newcommand{\contractprog}{\ensuremath{(\ekdoutput,\ekdstatenew) := \ekdprog(\ekdstateprev, \ekdinput)}}
\newcommand{\ekdcache}{\ensuremath{\mathsf{Cache}}}
\newcommand{\ekdbatch}{\ensuremath{\mathsf{Batch}}}

\newcommand{\enclaveparty}{\textcolor{party}{\ensuremath{\mathcal{E}}}}

\newcommand{\onrecv}{\textcolor{entry}{\bf On receive}\xspace}
\newcommand{\oninput}{\textcolor{entry}{\bf On input}\xspace}

\newcommand{\stringlitt}[1]{\textcolor{string}{\text{``#1''}}}

\newcommand{\msgok}{\stringlitt{ok}}
\newcommand{\msgcreate}{\stringlitt{create}}
\newcommand{\msgread}{\stringlitt{read}}
\newcommand{\msgwrite}{\stringlitt{write}}
\newcommand{\msgreceipt}{\stringlitt{receipt}}
\newcommand{\msgcheckreceipt}{\stringlitt{$\in$}}
\newcommand{\msgreject}{\stringlitt{reject}}
\newcommand{\msgquery}{\stringlitt{request}}

\newcommand{\msgcommitbatch}{\stringlitt{commit batch}}

\newcommand{\msgclaimoutput}{\stringlitt{claim output}}
\newcommand{\msginit}{\stringlitt{init}}
\newcommand{\msginstall}{\stringlitt{install}}
\newcommand{\msgresume}{\stringlitt{resume}}

\newcommand{\msginputkey}{\stringlitt{input key}}
\newcommand{\msgstatekey}{\stringlitt{state key}}
\newcommand{\msgoutputkey}{\stringlitt{output key}}
\newcommand{\msgcachemiss}{\stringlitt{cache miss}}
\newcommand{\msginterimoutput}{\stringlitt{atom-deliver}}
\newcommand{\msgfinaloutput}{\stringlitt{output}}

\newcommand{\fekiden}{\textcolor{party}{\ensuremath{\mathcal{F}_\text{\systemname}}}}
\newcommand{\fstorage}{\ensuremath{\textcolor{party}{\mathcal{F}_\text{blockchain}}}}
\newcommand{\funclink}{\ensuremath{\mathsf{succ}}}
\newcommand{\funcsgx}{\ensuremath{\textcolor{party}{\mathcal{G}_\text{att}}}}
\newcommand{\protoekiden}{\ensuremath{\mathbf{Prot}_\text{Ekiden}}}
\newcommand{\protoekidenfull}{\ensuremath{\mathbf{Prot}_\text{Ekiden}^\text{full}}}
\newcommand{\protowrapper}[1]{\widehat{#1}}

\newcommand{\pubkeyinput}{\ensuremath{\pk^{\text{in}}_{\ekdcid}}}
\newcommand{\seckeyinput}{\ensuremath{\sk^{\text{in}}_{\ekdcid}}}
\newcommand{\outputkey}{\ensuremath{\key^{\text{out}}_{\ekdcid}}}
\newcommand{\seckeycont}{\ensuremath{\key^\text{state}_{\ekdcid}}}

\newcommand{\pksgx}{\ensuremath{\pk_\text{\tee}}}
\newcommand{\sksgx}{\ensuremath{\sk_\text{\tee}}}
\newcommand{\sgxkeypair}{\ensuremath{(\pksgx,\sksgx)}}
\newcommand{\sgxsig}{\ensuremath{\Sigma_\text{\tee}}}
\newcommand{\sgxprog}{\ensuremath{\mathsf{prog}}}

\newcommand{\aesys}{\ensuremath{\mathcal{AE}}}
\newcommand{\sesys}{\ensuremath{\mathcal{SE}}}
\newcommand{\esk}{\ensuremath{\mathsf{esk}}}
\newcommand{\epk}{\ensuremath{\mathsf{epk}}}
\newcommand{\ssk}{\ensuremath{\mathsf{ssk}}}
\newcommand{\spk}{\ensuremath{\mathsf{spk}}}

\newcommand{\computenode}{\ensuremath{\mathsf{Comp}}\xspace}

\renewcommand{\pccomment}[1]{\textcolor{gray}{\scriptsize // #1}}

\newcommand{\protocol}[3][\columnwidth]{
	\begin{boxedminipage}[t]{#1}
		\begin{center}
		\scriptsize{\textbf{#2}}
		\end{center}
		\vspace{-\baselineskip}
		\procedure[mode=text, linenumbering, codesize=\scriptsize]{}{			
			#3
		}
	\end{boxedminipage}
}

\newcommand{\protocolsidebyside}[3]{
\begin{boxedminipage}[t]{\textwidth}
	\begin{center}
	\scriptsize{\textbf{#1}}
	\end{center}
	\vspace{-\baselineskip}
	\begin{pchstack}[center]
		\procedure[mode=text, linenumbering, codesize=\footnotesize]{}{
		    #2
		}
		\procedure[lnstart=21,mode=text, linenumbering, codesize=\footnotesize]{}{
			#3
		}
	\end{pchstack}
\end{boxedminipage}
}

\newcommand{\etal}{\textit{et al.}}
\newcommand{\tee}{TEE\xspace}
\newcommand{\tees}{TEEs\xspace}
\newcommand{\compactlist}[1]{
\begin{itemize}[noitemsep,topsep=2pt,parsep=0pt,partopsep=0pt]
    #1
\end{itemize}
}

\newcommand{\subparagraph}[1]{\vspace{2mm}\noindent\textbf{#1}} 
\begin{document}

\title{\systemname{}: A Platform for Confidentiality-Preserving, Trustworthy, and Performant Smart Contracts}

\IEEEoverridecommandlockouts
\IEEEaftertitletext{\centering This is an extended version of the EuroS\&P paper~\cite{confversion}.\vspace{4mm}}

\author{
  \IEEEauthorblockN{  Raymond Cheng\IEEEauthorrefmark{1}\IEEEauthorrefmark{4}
  Fan Zhang\IEEEauthorrefmark{2}
  Jernej Kos\IEEEauthorrefmark{4}
  Warren He\IEEEauthorrefmark{1}\IEEEauthorrefmark{4}
  Nicholas Hynes\IEEEauthorrefmark{1}\IEEEauthorrefmark{4}
  Noah Johnson\IEEEauthorrefmark{1}\IEEEauthorrefmark{4}
  }
  \IEEEauthorblockN{  Ari Juels\IEEEauthorrefmark{2}
  Andrew Miller\IEEEauthorrefmark{3}
  Dawn Song\IEEEauthorrefmark{1}\IEEEauthorrefmark{4}
  }
  \\
  \IEEEauthorblockA{
  \IEEEauthorrefmark{1}UC Berkeley
  \IEEEauthorrefmark{2}Cornell Tech
  \IEEEauthorrefmark{3}UIUC
  \IEEEauthorrefmark{4}Oasis Labs
  }
}

\maketitle

\todo{COMMENTS ON}

\begin{abstract}

Smart contracts are applications that execute on blockchains. Today they manage billions of dollars in value and motivate visionary plans for pervasive blockchain deployment. While smart contracts inherit the availability and other security assurances of blockchains, however, they are impeded by blockchains’ lack of {\em confidentiality} and poor {\em performance}. 

We present \systemname{}, a system that addresses these critical gaps by combining blockchains with Trusted Execution Environments (\tees). \systemname{} leverages a novel architecture that separates consensus from execution, enabling efficient \tee-backed \smartcontracts and high scalability. Our prototype (with Tendermint as the consensus layer) achieves example performance of 600x more throughput and 400x less latency at 1000x less cost than the Ethereum mainnet.

Another contribution of this paper is that we systematically identify and treat the pitfalls arising from harmonizing \tees and blockchains. Treated separately, both \tees and blockchains provide powerful guarantees, but hybridized, though, they engender new attacks. For example, in na\"i{ve} designs, privacy in TEE-backed contracts can be jeopardized by forgery of blocks, a seemingly unrelated attack vector. We believe the insights learned from \systemname{} will prove to be of broad importance in hybridized \tee-blockchain systems.

\end{abstract}

\section{Introduction}
\label{sec:introduction}

Smart contracts are protocols that digitally enforce agreements between or among distrusting parties. Typically executing on blockchains, they enforce trust through strong integrity assurance: Even the creator of a smart contract cannot feasibly modify its code or subvert its execution.
Smart contracts have been proposed to improve
applications across a range of industries, including finance, insurance, identity management, and supply chain management. 

Smart contracts inherit some undesirable blockchain properties. To enable validation of state transitions during consensus, blockchain data is public. Existing smart contract systems thus {\em lack confidentiality or privacy}: They cannot safely store or compute on sensitive data (e.g., auction bids, financial transactions). Blockchain consensus requirements also hamper smart contracts with {\em poor performance} in terms of computational power, storage capacity, and transaction throughput. 
Ethereum, the most popular decentralized smart contract platform, is used almost exclusively today for technically simple applications such as tokens, 
and can incur costs vastly (eight orders of magnitude) more than ordinary cloud-computing environments.   In short, the \emph{application complexity of smart contracts today is highly constrained}. Without critical performance and confidentiality improvements, smart contracts may fail to deliver on their transformative promise. 

Researchers have explored cryptographic solutions to these challenges, such as various zero-knowledge proof systems~\cite{kosba2016hawk} and secure multiparty computation~\cite{zyskind2015decentralizing}. However, these approaches have significant performance overhead and are only applicable to limited use cases with relatively simple computations. A more performant and general-purpose option is use of a {\em trusted execution environment} (\tee).

A \tee provides a fully isolated environment that prevents other software applications, the operating system, and the host owner from tampering with or even learning the state of an application running in the \tee. For example, Intel Software Guard eXtensions (SGX) provides an implementation of a \tee. The Keystone-enclave project~\cite{keystone} aims to provide an open-source \tee design.

A key observation driving our system design is that \tees and blockchains have complementary properties. On the one hand, a blockchain can guarantee strong availability and persistence of its state, whereas a \tee cannot guarantee availability (as the host can terminate \tees at its discretion), nor can it reliably access the network or persistent storage. On the flip side, a blockchain has very limited computation power, and must expose its entire state for public verification, whereas a \tee incurs minimal overhead compared with native computation, and offers verifiable computation with confidential state via remote attestation. Thus it appears appealing to build hybrid protocols that take advantage of both.

Harmonizing \tees with blockchains, though, is a challenge. Subtle pitfalls arise when the two are na\"{i}vely glued together. 

One such pitfall arises from a fundamental limitation of \tees: A malicious host can arbitrarily manipulate their scheduling and I/O. Consequently, \tees might terminate at any point, posing the risk and challenge of lost and/or conflicting state.
This problem is exacerbated by the fact that the so-called trusted timer in \tees (SGX, in particular) can in fact only provide a ``no-earlier-than'' notion of time, because a malicious host can also delay the clock read (a message transmitted over the bus).
Thus, while it's tempting to use a blockchain to checkpoint a \tee's state (e.g.~\cite{kaptchuk2017giving}), the lack of a reliable timer renders it tricky for a \tee to ascertain an up-to-date view of the blockchain. As we'll show later, na\"{i}ve state-checkpointing protocols open up rewinding attacks (\cref{sec:protocoldesignpricniples}). Another interesting and dangerous consequence is that seemingly unrelated attack vectors come into play. For example, the confidentiality of \tee-protected content could be jeopardized by integrity attacks against the blockchain: e.g., an attacker could circumvent a privacy budget enforced by a \tee by providing a forged blockchain to rewind its execution and sent it arbitrarily many queries. Other challenges include tolerating compromised \tees, supporting robust and consistent failover when \tees crash, and key management for enclaves. We systematically identify and treat each of these pitfalls in this paper.

Following the above design principles,
we present \systemname{}, a system for highly performant and \smartcontracts. 
To the best of our knowledge, \systemname{} is the first \smartcontract system capable of thousands of transactions per second. 
The key to this achievement is a secure and principled combination of blockchains and trusted hardware.
\systemname{} combines any desired underlying blockchain system (permissioned or permissionless) with \tee-based execution. Anchored in a formal security model expressed as a cryptographic ideal functionality~\cite{ucCanetti}, \systemname{}'s principled design supports rigorous analysis of its security properties. 

\systemname{} adopts an architecture where \emph{computation} is separated from \emph{consensus}.
\systemname{} uses \emph{compute nodes} to perform smart contract computation over private data off chain in \tees, then attest to their correct execution on chain. The underlying blockchain is maintained by  \emph{consensus nodes}, which need not use trusted hardware. \systemname{} is agnostic to consensus-layer mechanics, only requiring a blockchain capable of validating remote attestations from compute nodes. \systemname{} can thus scale consensus and compute nodes independently according to performance and security needs.

By operating compute nodes in \tees, \systemname{} imposes minimal performance overhead relative to an ordinary (e.g., cloud) computing environment.
In this way, we avoid the computational burden and latency of on-chain execution. 
\tee-based computation in \systemname{} provides confidentiality, enabling efficient use of powerful cryptographic primitives that a \tee is known to emulate, such as functional encryption~\cite{fisch2017iron} and black-box obfuscation~\cite{nayak2017hop}, and also provides a trustworthy source of randomness, a major acknowledged difficulty in blockchain systems~\cite{bunz2017proofsof}.
	
To address the availability and network security limitations of \tees, 
\systemname{} supports on-chain checkpointing and (optional) storage of contract state. 
\systemname{} thereby supports safe interaction among long-lived smart contracts across different trust domains. 
To address potential TEE failures, such as side channel attacks, we propose mitigations
to preserve integrity and limit data leakage (\Cref{sec:toleratehostfailures}).
Assuming blockchain integrity, users need not trust smart contract creators, miners, node operators or any other entity for liveness, persistence, confidentiality, or correctness.  \systemname{} thus enables self-sustaining services that can outlive any single node, user, or development effort.\footnote{Our system name \systemname{} refers to this property.  ``Ekiden'' is a Japanese term for a long-distance relay running race.}

\noindent\textbf{Technical challenges and contributions.}
Our work on \systemname{} addresses several key technical challenges:
\begin{itemize}
\item{\em Formal security modeling:} While intuitively clear, the desired and achievable security properties required for \systemname{} are challenging to define formally. 
We express the full range of security requirements of \systemname{} in terms of an ideal functionality $\fekiden$.
We outline a security proof in the Universal Composability (UC) framework that shows that the \systemname{} protocol matches $\fekiden$ under concurrent composition. 
\item{\em A principled approach for hybridized \tee-blockchain systems:}
We systematically enumerate the fundamental pitfalls arising from fusing blockchains and \tees and offer general techniques for overcoming them. 
Further, we show that by appealing to cryptographic ideal functionalities, these techniques can be applied in a principled, provably secure, and performant way that we believe can be generalized to a broad range of hybridized \tee-blockchain systems. 

\item{\em Performance:} The blockchain is likely to be a performance bottleneck of a \tee-blockchain hybrid system. We provide optimization that minimize the use of blockchain without degrading security: We show that they realize the same $\fekiden$ functionality as the unoptimized protocol. \end{itemize}

\noindent\textbf{Evaluation.}
We evaluate the performance of \systemname{} on a suite of applications that exercise the full range of system resources 
and demonstrate how \systemname{} enables application deployment that would otherwise be impractical
due to privacy and/or performance concerns.
They include a machine learning framework, within which we implement medical-diagnosis and credit-scoring applications,
a smart building thermal model, and a poker game.
We also port an Ethereum Virtual Machine implementation to \systemname{}, 
so that existing contracts (e.g., written in Solidity),
such as Cryptokitties~\cite{cryptokitties} and the ERC20 token,
can run in our framework as well. 
We report on development effort, showing that the programming model in \systemname{} lends itself 
to simple and intuitive application development. 
Contracts in \systemname{} process transactions 2--3 orders of magnitude
both faster and higher throughput over Ethereum.
Our performance optimizations also greatly compress the amount of data
stored on the blockchain, yielding a 2--4 order of magnitude improvement over the baseline. (The advantage is greater for read-write operations on contracts with large state, such as our token contract.)

\section{Background}
\label{sec:background}

\paragraph{Smart Contracts and Blockchains}

Blockchain-based smart contracts are programs executed by a network of participants
who reach agreement on the programs' state.
Existing smart contract systems replicate data and computation
on all nodes in the system.
so that individual node can verify correct execution of the contract.
Full replication on all nodes provides a high level of fault tolerance and availability.
Smart contract systems such as Ethereum~\cite{ethereum} has 
demonstrated their utility across a range of applications.

However, several critical limitations
impede wider adoption of current smart contract systems.
First, on-chain computation of fully replicated smart contracts is inherently expensive. 
For example in August 2017, it cost \$26.55 to add 2 numbers together one million times
in an Ethereum smart contract~\cite{ethereum},
a cost roughly 8 orders of magnitude higher than in AWS EC2~\cite{ethereum-costs}.
Furthermore, current systems offer no privacy guarantees.
Users are identified by pseudonyms.
As numerous studies have shown~\cite{reid2013analysis,meiklejohn2013fistful,moser2017price,ron2013quantitative}, pseudonymity provides only weak privacy protection.
Moreover, {\em contract state and user input must be public} in order for miners to verify correct computation.
Lack of privacy fundamentally restricts the scope of applications of smart contracts.

\paragraph{Trusted Hardware with Attestation}

A key building block of \systemname{} is a trusted execution environment (\tee) that protects the confidentiality and integrity of computations, and can issue proofs, known as {\em attestations}, of computation correctness.  \systemname{} is implemented with Intel SGX~\cite{Anati2013,Hoekstra2013,McKeen2013}, a specific \tee technology, but we emphasize that it may use any comparable \tee with attestation capabilities, such as the ongoing effort Keystone-enclave~\cite{keystone} aiming to realize open-source secure hardware enclave.
We now offer brief background on \tees, with a focus on Intel SGX.

Intel SGX provides a CPU-based implementation of \tees---known as \emph{enclaves} in SGX---for general-purpose computation.
A host can instantiate multiple \tees, which are not only isolated from each other, but also from the host.
Code running inside a \tee has a protected address space.
When data from a \tee moves off the processor to memory, it is transparently encrypted
with keys only available to the processor.
Thus the operating system, hypervisor, and other users cannot access the enclave's memory.
The SGX memory encryption engine also guarantees data integrity and prevents memory replay attacks~\cite{gueron2016memory}.
Intel SGX supports attested execution, i.e., it is able to prove the correct execution of a program, by issuing a {\em remote attestation}, a digital signature, using a private key known only to the hardware, over the program and an execution output. Remote attestation also allows remote users to establish encrypted and authenticated channels to an enclave~\cite{Anati2013}. Assuming trust in the hardware, and Intel, which authenticates attestation keys, it is infeasible for any entity other than an SGX platform to generate any attestation, i.e., attestations are existentially unforgeable.

However, attested execution realized by trusted hardware isn't perfect.
For example, SGX alone cannot guarantee availability: a malicious host can terminate enclaves or drop messages arbitrarily.
Even an honest host could accidentally lose state (e.g. when power cycles).
The weak availability of SGX poses a fundamental challenge to the design of \systemname{}. 
Also, the current SGX implementation is vulnerable to side-channel attacks~\cite{DBLP:conf/sp/XuCP15,SGXSpectre}.
\systemname{} is compatible with existing defenses~\cite{bernstein2012security,nayak2017hop,Liu:2015:OPF:2867539.2867660,sgp,rane2016raccoon}. We discuss side-channel resistance in \cref{sec:toleratehostfailures}.

\section{Technical Challenges in \tee-blockchain hybrid systems}
\label{sec:protocoldesignpricniples}

Before diving into the specifics of \systemname{}, we first describe and address the fundamental pitfalls that arise when harmonizing \tees and blockchains. The solutions serve as building blocks of the \systemname{} protocol, and we believe the insights learned from \systemname{} will prove to be of broad importance in hybridized \tee-blockchain systems.

\subsection{Tolerating \tee failures}
\label{sec:toleratehostfailures}

Although designed to execute general purpose programs, trusted hardware is not a panacea. Here we analyze the limitations of \tees and their impact on \tee-blockchain hybrid protocols.

\paragraph{Availability failures}

Trusted hardware in general cannot ensure availability. In the case of SGX, a malicious host can terminate enclaves, and even an honest host could lose enclaves in a power cycle. A \tee-blockchain system must tolerate such host failures, ensuring that crashed \tees can at most delay execution.

Our high-level approach is to treat \tees as expendable and interchangeable, relying on the blockchain to resolve any conflicts resulting from concurrency. To ensure that any particular \tee is easily replaced, \tees are \emph{stateless}, and any persistent state is stored by the blockchain. We discuss later how \tees can also keep soft state across invocations as a performance optimization, but we emphasize that the techniques in \systemname{} ensure that losing such state at any point does not affect security.

\paragraph{Side channels}
Although \tees aim to protect confidentiality, recent work has uncovered data leakage via side-channel attacks.
Existing defenses are generally application- and attack-specific 
(e.g., crypto libraries avoid certain data-dependent operations~\cite{bernstein2012security});
generalizing such protections remains challenging.
Thus, \systemname{} largely defers protections to the application developer.

Even though there is perhaps no definitive and practical panacea to all side-channel attacks, it is still desirable to limit the impact of compromised \tees and provide graceful degradation in the face of small-scale compromise.
Our approach is to compartmentalize both spatially and temporally.
We design critical components in \systemname{}, such as the key manager, against a strong adversarial model, allowing an attacker to break the confidentiality of a small fraction of \tees, and limit the access to the key manager from other components. We also employ proactive key rotation~\cite{Herzberg95proactive} to confine the purview of a leaked key. Key management is fundamental to the availability of a \tee-blockchain system, as discussed below.

\paragraph{Timer failures}
\tees in general lack trusted time sources.
In the case of SGX, although a trusted relative timer is available,
the communication between enclaves and the timer (provided by an off-CPU component) can be delayed by the OS~\cite{sgxDelayedTrustedTime,IntelSGXPS}.
Moreover server-grade Intel CPUs offer no support for SGX timers at the time of writing. Thus a \tee-blockchain hybrid protocol must minimize reliance on the \tee timer.

Our approach is to design protocols that do not require \tees to have a current view of a blockchain. Specifically, instead of requiring a \tee to distinguish stale state from current state (without a synchronized clock, there is no definitive countermeasure to a network adversary delaying messages from the blockchain), our techniques rely on the blockchain to proactively reject any update based on a stale input state (a hash of which is included in the update). 

The missing timer also makes it hard for \tees
to verify that an item has been persisted in the blockchain, i.e. to establish ``proofs of publication,'' as coined by~\cite{kaptchuk2017giving}. However \cite{kaptchuk2017giving} doesn't consider threats caused by lack of trustworthy time in \tees---e.g., injection of old, fake, easily minable blocks---that are critical in PoW-based blockchains. One of our contributions is a general, time-based proof-of-publication protocol
that is secure against network adversary delaying clock read, as we now briefly explain.

\subsection{Proof of Publication for PoW blockchains}
\label{subsec:PoPub}
In order to leverage blockchains as persistent storage, a \tee must be able to efficiently verify that an item has been stored in the blockchain.
For permissioned blockchains,
such a proof can consist of signatures from a quorum of consensus nodes.
To establish proofs of publication for PoW-based blockchains,  \tees must be able to validate new blocks.
As noted in~\cite{fairnessChoudhuri}, a trusted timer is needed to defend 
against an adversary isolating an enclave and presenting an invalid subchain.
Unfortunately, timing sources over secure channels (e.g. SGX timers) cannot guarantee a bounded response time, as discussed above.
To work around this limitation, we leverage the confidentiality of \tees so that an attacker
delaying a timer's responses cannot prevent an enclave from successfully verifying blockchain contents.
Our solution can even work without SGX timers given trust in, e.g. TLS-enabled NTP servers.
Due to lack of space, we relegate our proof-of-publication protocol for PoW blockchains to \cref{sec:proof_pub}.

\subsection{Key management in \tees}

A fundamental limitation of using a blockchain to persist \tee state is the lack of confidentiality. We showed previously how to avoid this problem by encryption. This, however, leads to another problem: how can one persist the encryption keys?

Generally the method is to replicate keys across multiple \tees.
However, the flip side is the challenge of minimizing the key exfiltration risk in the face of confidentiality breach (e.g. via side-channel attacks).
There is in general a fundamental tension between exposure risk and availability:
A higher replication factor means not only better resiliency to state loss, but also a larger attack surface.
Therefore the tradeoff and achievable properties would depend on the threat model.

Since there is perhaps no definitive and practical full-system side-channel mitigation,
our approach is to design the key manager against a stronger adversarial model where the attacker is allowed to break the confidentiality of a small fraction of \tees, and limit the access from other components. 
We outline the key management protocol in \cref{sec:keymanagement}.

\subsection{Atomic delivery of execution results}
\label{sec:atomic-overview}

In blockchain systems, ensuring the atomicity of executions, namely either both executions $e_1, e_2$ finish or none of them, has been a fundamental problem, as exemplified by work on atomic cross-chain swaps~\cite{Tesseract}. A similar but more complicated problem arises in \tee-blockchain hybridization.

For a general stateful \tee-blockchain protocol,
\tee execution yields two messages:
$m_1$, which delivers the output to the caller, and $m_2$,
which delivers the state update to the blockchain,
both via adversarial channels.
We emphasize that it is critical to enforce \textbf{atomic delivery} of the two messages, i.e.
both $m_1$ and $m_2$ are delivered or the system has become permanently unavailable.
$m_1$ is delivered when the caller receives it.
The new state $m_2$ is delivered once accepted by the blockchain.
Rejected state update are not considered delivered. 

To see the necessity of atomic delivery,
consider possible attacks when it's violated,
i.e., when only one of the two messages is delivered.
First, if only the output $m_1$ is delivered,
a \emph{rewind attack} becomes possible.
Since \tee cannot tell whether an input state is fresh,
an attacker can provide stale states to resume a \tee's execution from an old state.
This enables grinding attacks against randomized \tee programs. An attacker may repeatedly rewind until receiving the desired output.
Another example is that rewinding could defeat budget-based privacy protection, such as differential privacy.
On the other hand, if only the state update $m_2$ is delivered,
the user risks permanent loss of the output,
as it might be impossible to reproduce the same output with the updated state.

We specify the atomic delivery protocol in \cref{sec:atomic}.

\section{Overview of \systemname{}}
\label{sec:overview}

In this section, we provide an overview of the design and security properties of \systemname{}.

\subsection{Motivation}
As an example to motivate our work, consider a credit scoring application---an example we implement and report on in~\cref{sec:impl:programming}.
Credit scores are widely used by lenders, insurers, and others to evaluate the creditworthiness of consumers.
Despite its considerable revenue (\$10.8B in 2017~\cite{creditbureau-stats}), the credit reporting industry in the U.S. is concentrated among a handful of credit bureaus~\cite{creditbureau-stats}. Such centralization creates large single points of failure and other problems, as highlighted by a recent data breach affecting nearly half the US population~\cite{hack-equifax}.

Blockchain-based decentralized credit scoring is thus an attractive and popular alternative. Bloom~\cite{Bloom}, for example, is a startup offering a credit scoring system on Ethereum. Their scheme, however, only supports a static credit scoring algorithm that omits important private data and cannot support predictive modeling. Such applications are bedeviled by two critical limitations of current smart contract systems: (1) A lack of \emph{data confidentiality} needed to protect sensitive consumer records (e.g., loan-service history for credit scoring) and the proprietary prediction models derived from them and (2) A failure to achieve the \emph{high performance} needed to handle global workloads. 

To support large-scale, privacy-sensitive applications like credit scoring,
it is essential to meet these two requirements while preserving the \emph{integrity}
and \emph{availability} offered by blockchains---all without requiring a trusted third party.
\systemname{} offers a confidential, trustworthy, and performant platform that achieves precisely
this goal for smart contract execution.

\begin{figure}[t!]
  \centering
  \includegraphics[width=\columnwidth]{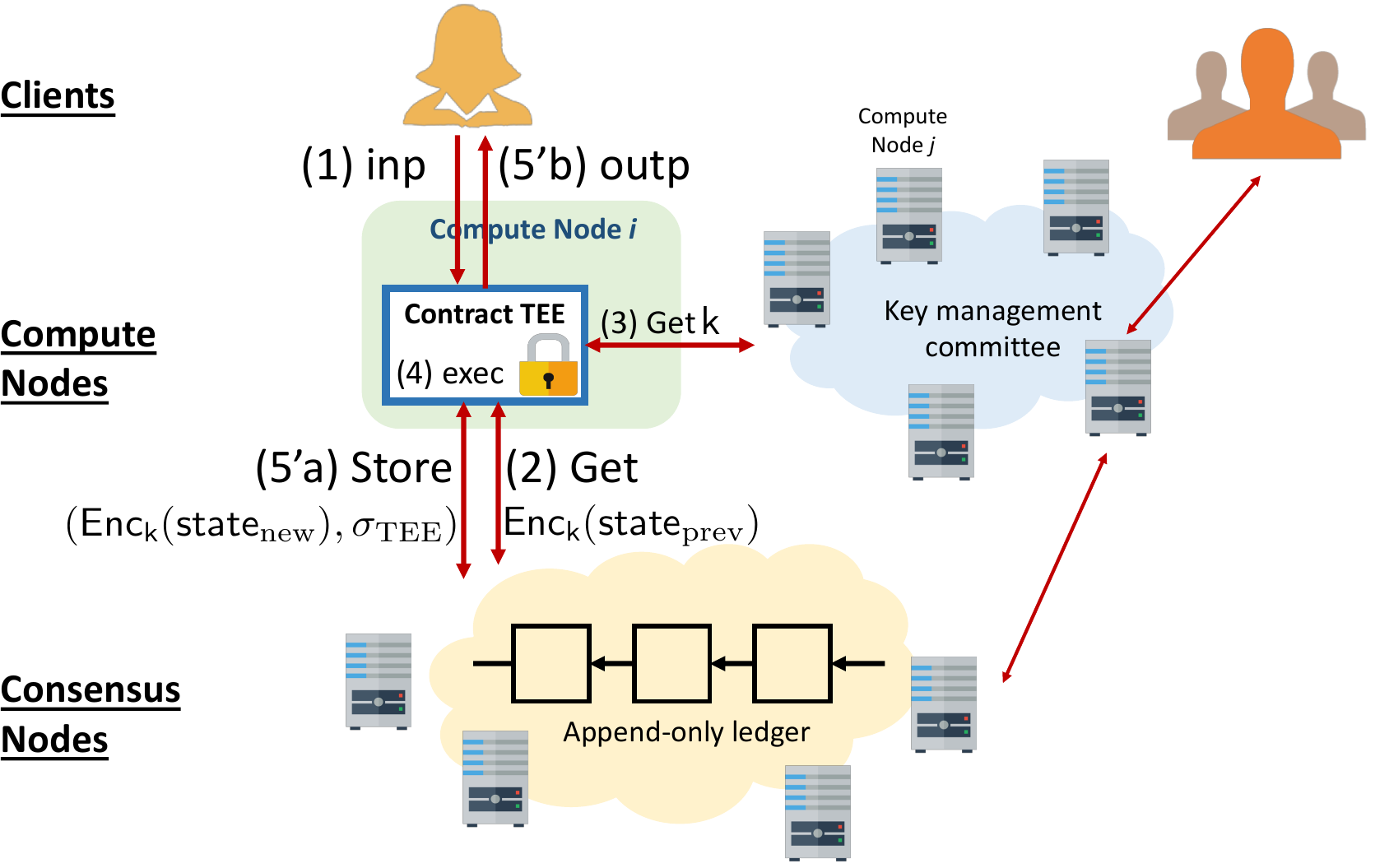}
  \caption{
    Overview of \systemname{} architecture and workflow.
    Clients send inputs to \smartcontracts, which are executed within a \tee at any compute node. The blockchain stores encrypted contract state. See \Cref{subsec:arch} for an overview.}
  \label{fig:workflow}
\end{figure}

\subsection{\systemname{} Overview}
\label{subsec:arch}

Conceptually, \systemname{} realizes a secure execution environment for rich user-defined smart contracts. An \systemname{} contract is a deterministic stateful program. Without loss of generality, we assume contract programs take the form $\contractprog$, ingesting as input a previous state $\ekdstateprev$ and a client's input $\ekdinput$, and generating an output $\ekdoutput$ and new state $\ekdstatenew$.

Once deployed on \systemname{}, smart contracts are endowed with strong confidentiality, integrity and availability guarantees. 
\systemname{} achieves these properties with a hybrid architecture combining trusted hardware and the blockchain. 
\Cref{fig:workflow} depicts the architecture of \systemname{} and a workflow of \systemname{} smart contracts. As it shows, there are three types of entities in \systemname{}: clients, compute nodes and consensus nodes.

\squishlist{}
    \item \textbf{Clients} are end users of smart contracts. In \systemname{}, a client can create contracts or execute existing ones with secret input. In either case, clients delegate computation to compute nodes (discussed below). We expect clients to be lightweight, allowing both mobile and web applications to interact with contracts.
    \item     \textbf{Compute nodes} process requests from clients by running the contract in a contract \tee and generating attestations proving the correctness of state updates.  Anyone with a \tee-enabled platform can participate as a compute node, contributing to the liveness and scalability of the system. A quorum of compute nodes form a key management committee and run a distributed protocol to manage keys used by contract \tees. A contract \tee reaches out to the key management committee to create or retrieve keys. We defer details of key management to \Cref{sec:keymanagement}.
    \item \textbf{Consensus nodes} maintain a distributed append-only ledger, i.e. a blockchain, by running a consensus protocol. Contract state and attestations are persisted on this blockchain. Consensus nodes are responsible for checking the validity of state updates using TEE attestations, as we discuss below.
\squishend{}

\subsection{Workflow} We now sketch the contract creation and request execution workflow, providing further details on \Cref{fig:workflow}. The detailed formal protocol is presented in \cref{sec:protocoldetails}. 

For simplicity, we assume a client has a priority list of compute nodes to use. In practice, a coordinator can be employed to facilitate compute node discovery and load balancing. We denote a client as $\ekdparty$ and a compute node as $\computenode$.

\paragraph{Contract creation}

When creating a contract, $\ekdparty$ sends a piece of contract code $\ekdprog$ to \computenode. \computenode loads $\ekdprog$ into a \tee (called contract \tee hereafter), and starts the initialization.
The contract \tee creates a fresh contract id $\ekdcid$, obtains fresh $(\pubkeyinput, \seckeyinput)$ pair and $\seckeycont$ from the key management committee and generates an encrypted initial state $\enc(\seckeycont, \vec{0})$ and an attestation $\ekdatt$, proving the correctness of initialization and that $\pubkeyinput$ is the corresponding public key for contract $\ekdcid$.
Finally, \computenode obtains a proof of the correctness of $\ekdatt$ by contacting the attestation service (detailed below); this proof and $\ekdatt$ are bundled into a ``certified'' attestation $\sigatt$.
\computenode then sends $(\ekdprog, \pubkeyinput, \enc(\seckeycont, \vec{0}), \sigatt)$ to consensus nodes. The full protocol for contract creation is specified in the $\msgcreate$ call of $\protoekiden$ (\cref{fig:protocolekiden}).
Consensus nodes verify $\sigatt$ before accepting $\ekdprog$, the encrypted initial state, and $\pubkeyinput$ as valid and placing it on the blockchain.

\paragraph{Request execution} The steps of request execution illustrated in \cref{fig:workflow} are as follows:

\squishdesc{}
    \item[(1)] To initiate the process of executing a contract $\ekdcid$ with input $\ekdinput$, $\ekdparty$ first obtains $\pubkeyinput$ associated with the contract $\ekdcid$ from the blockchain, computes $\ekdinputct = \enc(\pubkeyinput, \ekdinput)$ and sends to \computenode a message $(\ekdcid, \ekdinputct)$, as specified in Lines 8-11 of $\protoekiden$.     \item[(2)] \computenode retrieves the contract code and the encrypted previous state $\ekdstatect=\ekdstateprevct$ of contract $\ekdcid$, from the blockchain, and loads $\ekdstatect$ and $\ekdinputct$ into a \tee and starts the execution, as specified in Line 30-33 of $\protoekiden$. 
    \item[(3-4)] From the key management committee, the contract \tee obtains $\seckeycont$ and $\seckeyinput$,
    with which it decrypts $\ekdstatect$ and $\ekdinputct$ and executes, generating an output $\ekdoutput$, a new encrypted state $\ekdstatect'=\ekdstatenewct$, and an signature $\sigatt$ proving correct computation, as specified in Line 7-13 of the \tee Wrapper (\cref{fig:enclavewrapper}). 
    \item[(5a, 5b)] Finally, \computenode and $\ekdparty$ conduct an atomic delivery protocol which delivers $\ekdoutput$ to $\ekdparty$ and $(\ekdstatect',\sigatt)$ to the consensus nodes.
    We defer the detail of atomic delivery to \Cref{sec:atomic}.
    Briefly, Step 5a and Step 5b in \cref{fig:workflow} are executed atomically, i.e.
    $\ekdoutput$ is revealed to $\ekdparty$ if and only if $(\ekdstatect',\sigatt)$ is accepted by consensus nodes.
    Consensus nodes verify $\sigatt$ before accepting the new state as valid and placing it on the blockchain.
\squishdescend{}

A key distinction between \systemname{} and existing smart contract platforms (e.g. Ethereum~\cite{ethereum}) is \systemname{} decouples request execution from consensus.
In Ethereum, request execution is replicated by all nodes in the network to reach consensus, rendering the entire network as slow as a single node. Whereas in \systemname{}, request is only executed by $K$ compute nodes for some small $K$ (e.g. in \Cref{fig:workflow}, we set $K=1$) and consensus nodes just verify $K$ proofs of correct execution without repeating the execution.

In our implementation, a proof of correct execution takes the form of a signature $\sigatt$. Specifically, a compute node \computenode obtains $\sigatt$ as follows.
Suppose the execution on \computenode results in an output $\ekdstatect'$ and an attestation $\ekdatt$ (a signature~\cite{epid} over the contract code and $\ekdstatect'$).
\computenode then sends $\ekdatt$ to the Intel Attestation Service (IAS), which verifies $\ekdatt$ and replies with $\sigatt=(b, \ekdatt, \sigias)$, where $b\in\bin$ indicates the validity of $\ekdatt$ and $\sigias$ is a signature over $b$ and $\ekdatt$ by IAS. $\sigatt$ is then submitted to consensus node as a proof of correctness for $\ekdstatect'$.
As $\sigatt$ is just a signature, consensus nodes need neither trusted hardware nor to contact the IAS to verify it.

\subsection{\systemname{} Security Goals}

Here we summarize the security goals of \systemname{}. Briefly,
\systemname{} aims to support execution of general-purpose contracts while enforcing the following security properties: 

\squishdesc{}

\item [Correct execution:]
Contract state transitions reflect correct execution of contract code on given state and inputs.

\item [Consistency:]
At any time, the blockchain stores a single sequence of state transitions
consistent with the view of each compute node.

\item [Secrecy:]
During a period without any \tee breach, \systemname{} guarantees that contract state and inputs from honest clients are kept secret from all other parties. Additionally, \systemname{} is resilient to some key-manager \tees being breached.

\item [Graceful confidentiality degradation:]
Should a confidentiality breach occur in a computation node (as opposed to a key-manager node), \systemname{} provides forward secrecy and reasonable isolation from the affected \tees. Specifically, suppose a confidentiality breach happens at $t$. The attacker can at most access the history up to $t - \Delta$ where $\Delta$ is a system parameter. Moreover, a compromised \tee can only affect a subset of contracts.

\squishdescend{}

\vspace{1mm}
\noindent\textbf{Non-goals:}
\systemname{} does \emph{not} prevent contract-level leakage (e.g. through covert channels, bugs or side channels).
Thus contract developers are responsible for ensuring that no secret is revealed through public output, and that the contract is free of bugs and side channels. We discuss supported mitigation in \cref{subsub:application-levelleakage}.

\subsection{Assumptions and Threat Model}
\label{subsec:threatmodel}

\paragraph{\tee}
Recent work demonstrates that the confidentiality of SGX enclaves may be compromised via side-channel attacks.
In light of this threat, we assume the adversary can compromise the confidentiality of a small fraction of \tees. As noted above, the impact depends on whether the breaches affect key-manager or computation nodes.
We assume that \tee hardware is otherwise correctly implemented and securely manufactured.

\paragraph{Blockchain}
\systemname is designed to be agnostic to the underlying consensus protocol.
It can be deployed atop any blockchain implementation
as long as the requirements specified below are met.

We assume the blockchain will perform prescribed computation correctly and is always available. In particular, \systemname{} relies on consensus nodes to verify attestations. We further assume the blockchain provides an efficient way to construct proofs of item inclusion on the blockchain, i.e., proofs of publication, as discussed in~\cref{subsec:PoPub}.

\paragraph{Threat Model}
All parties in the system must trust \systemname{} and \tee.
We assume the adversary can control the operating system and the network stack of all but one compute nodes. On controlled nodes, the adversary can reorder messages and schedule processes arbitrarily. We assume the attacker can compromise the confidentiality of a small fraction (e.g. $f\%$) of \tees. The adversary observes global network traffic and may reorder and delay messages arbitrarily.

The adversary may corrupt any number of clients.
Clients need not execute contracts themselves and do not require trusted hardware. 
We assume honest clients trust their own code and platform, but not other clients.
Each contract has an explicit policy dictating how data is processed and requests are serviced. \systemname{} does not (and cannot reasonably) prevent contracts from leaking secrets intentionally
or unintentionally through software bugs.

\section{Building blocks}

Before diving to protocol details, we first present key building blocks of the \systemname{} protocol, addressing the general technical challenges in \tee-blockchain systems, as reviewed in \cref{sec:protocoldesignpricniples}.

\subsection{Proof of Publication}
\label{sec:proof_pub}

We now present a proof of publication protocol for permissionless blockchains. Please refer to \cref{subsec:PoPub} for background and motivation. A proof of publication is an interactive proof between a verifier $\enclaveparty$, in the form of a contract \tee, and a untrusted prover $\ekdparty$. The high level idea is to only give $\ekdparty$ a limited amount of time to publish the message in a block within a subchain of sufficient difficulty so that an adversary cannot feasibly forge it. The protocol is formally specified in~\onlyinfullversion{fig:proofofpublication}. We give text description below so the formal specification is not required for understanding.

$\enclaveparty$ stores a recent checkpoint block $CB$ from the blockchain, from which a difficulty $\delta(CB)$, e.g. the number of leading zeroes in the block nonce, can be calculated. $\enclaveparty$ will emit an (attested) version of $CB$ to any requesting client, enabling the client to verify $CB$'s freshness.
Given a valid recent $CB$, $\enclaveparty$ can verify new blocks based on $\delta(CB)$, assuming the difficulty is relatively stationary. (For simplicity in our analysis here, we assume constant difficulty, but our analysis can be extended under an assumption of bounded difficulty variations.)

To initiate publication of $m$, $\enclaveparty$ calls the timer to get a timestamp $t_1$. As discussed, $\enclaveparty$ may receive $t_1$ after a delay. 
After receiving $t_1$ (maybe at a time later than $t_1$), $\enclaveparty$ generates a random nonce $r$ and requires the prover to publish $(m,r)$. Upon receiving a proof $\pi_{(m,r)}$ (a subchain containing $(m,r)$) from $\ekdparty$, $\enclaveparty$ calls the timer again for $t_2$.
Let $n_c$ to be the number of confirmations in $(m,r)$, $\tau$ be the expected block interval (an invariant of the blockchain), and $\epsilon$ be a multiplicative {\em slack} factor that accounts for  variation in the time to generate blocks, which is a stochastic process. E.g., $\epsilon=1.5$ means that production of $\pi_{(m,r)}$ is allowed to be up to $1.5$ times slower than expected on the main chain. $\enclaveparty$ accepts $\pi_{(m,r)}$ only if $t_2 - t_1 < n_c\times\tau\times\epsilon$.

Setting $\epsilon$ to a high value reduces the probability of false rejections (i.e., rejecting proofs from an honest $\ekdparty$ when the main chain growth was unluckily slow during some timeframe). However, a high $\epsilon$ also increases the possibility of false acceptance, i.e. accepting a forged subchain. For any $\epsilon > 1$, it is possible to require a
large enough $n_c$ so that the probability of a successful attack becomes negligible. However, a large $n_c$ means that an honest $\ekdparty$ needs to wait for a long time before $\ekdparty$ can obtain the output, may affecting the user experience.

For example, for a powerful attacker with $25\%$ hash power (roughly the largest mining pool known to exist in Bitcoin and Ethereum at the time of writing), setting $n_c=80$ and $\epsilon=1.6$ means the attacker needs an expected $2^{112}$ hashes to forge a proof of publication\footnote{as the time of writing, it takes roughly $2^{73}$ hashes to mine a Bitcoin block.}, while an honest proof will be rejected with probability $2^{-19}$. Similar block-synchronization techniques and analysis are used in the recently proposed Tesseract \tee-based cryptocurrency exchange~\cite{Tesseract}.

It is easy to see that delaying the timer's responses does not give the attacker more time than $t_2 - t_1$. Delaying timestamp $t_1$ shrinks this apparent interval of time, disadvantaging the attacker.
$\enclaveparty$'s checkpoint block can be updated with the same protocol, by publishing an empty message. Note that once a message is successfully published by a \tee, other \tees can obtain the proof via secure channels established by attestations, saving the cost of repeating the protocol.

\subsection{Key Management}
\label{sec:keymanagement}

Each \systemname{} contract is associated with a set of keys, including a symmetric key for state encryption and a key pair to encrypt client input. Here we discuss the generation, distribution, and rotation of these keys.

\subsubsection{Adversarial model}

We consider a adversary that can break the confidentiality, e.g., via side-channel attacks, of some fraction (e.g. $f\%$) of the \tees. The exact value of $f$ depends on the deployment and enrollment model. $f$ can be a very low value if enrollment is limited to well-managed nodes, e.g., ones hosted by capable and reputable organizations. But when deployed in a more open environment, $f$ needs to be reasonably high. We assume the participating hosts have (at least partially) Sybil-resistant identities. One way to achieve this is to require a security deposit to join the protocol.

In addition, we assume there are sufficiently many (e.g. more than $2f\%$ of) participants online at any time so that the availability of keys are retained. In practice, participation can be motivated by economic rewards and penalties. We leave the incentive design for future work. 

\subsubsection{Desired properties}

Since decryption keys are eventually revealed to a contract \tee, which itself may also be compromised, actively used keys (i.e. hot keys) must be short-live, derived from a less-exposed long-term master secret. Ideally, a key management protocol should satisfy the following properties:

\vspace{2mm}
\squishlist{}
\item \textit{Confidentiality}: The adversary (within our model) cannot exfiltrate the long-term master key.
\item \textit{Availability}: An honest contract \tee can always access decryption keys.
\item \textit{Forward secrecy}: If a short-term key is compromised at time $t$, it cannot be used to decrypt messages encrypted before $t - \Delta$, for some system parameter $\Delta$.
\squishlistend{}
\vspace{2mm}

\subsubsection{Preliminaries}
Below we outline a key management protocol that satisfies the above requirements. We first review the building blocks, including distributed key generation (DKG) protocols and distributed pseudo-random  functions (PRFs).

\paragraph{Distributed Key Generation (DKG)}
A DKG protocol (e.g.~\cite{gennaro1999secure}) allows a set of $N$ parties to generate unbiased, random keys. The outcome of a run of a DKG protocol is a secret $s$, but shared among parties using a secret-sharing scheme (typically Shamir's).

\paragraph{Distributed PRF}

Informally, a PRF is a collection of functions $\mathcal{F}=\set{f_s}_{s\in S}$, such that for a random index $s\sample S$, $f_s(\cdot)$ is indistinguishable from a random function.

Naor \etal~\cite{naor1999distributed} introduce distributed PRFs, which are such that parties with shares of $s$ can evaluate $f_s(\cdot)$ without reconstructing $s$. Specifically, let $G$ be a Schnorr group and $g$ be a generator. Let $\hash:\bin^* \to G$ be a hash function, \cite{naor1999distributed} shows that $f_s(x) = \hash(x) ^ s$ is a family of PRF.

Suppose $s$ is shared among parties using a $(k, n)$-secret sharing scheme. To evaluate $f_s(x)$, party $i$ simply computes and outputs $y_{i} = \hash(x) ^ {s_i}$, computed with its share $s_i$. After collecting at least $k+1$ of $\set{y_{i}}$, one can derive $f_s(x)$ by polynomial interpolation in the exponent:
\begin{equation*}
    f_s(x) = \hash(x) ^ {S} = \hash(x) ^ {\sum_{i \in A} S_i \lambda_i} = \prod_{i\in A} y_{i}^{\lambda_i}
\end{equation*}
where $\lambda_i$ are Lagrange coefficients $\lambda_i = \prod_{j\ne i} \frac{-j}{i-j}$.

\subsubsection{Protocol}

\paragraph{Key management committees and long-term keys}

Assuming Sybil-resistant identities, we can sample $N$ nodes from the participants to form a key management committee (KMC). $N$ is a system parameter. When initializing a contract $c$, KMC runs the DKG protocol to generate a long term key $\key_c$, so that $\key_c$ is secret-shared among KMC members using a $(\ceil{fN}, N)$-secret sharing scheme. Previous work on proactive secret sharing (e.g.~\cite{Herzberg95proactive, schultz2010mpss}) can be used to periodically rotate the committee without changing the secret. \cite{schultz2010mpss} also allows a committee to be dynamically expanded.

\paragraph{Generating short-term keys}

Suppose short-term keys expire every epoch.
To get the short-term key for contract $c$ at epoch $t$, a compute node $\computenode$ first establishes secure channels and authenticates itself with members in KMC. Once verified that $\computenode$ is indeed executing $c$, each KMC member $i$ computes $\key_{c,t,i} = \hash(t) ^ {\key_c^i}$ and sends $\key_{c,t,i}$ to $\computenode$.
After collecting $f+1$ outcomes from $A\subseteq \text{KMC}$, $\computenode$ can construct the short-term key for epoch $t$ by $\key_{c,t} = \prod_{i\in A} \key_{c,t,i}^{\lambda_i}$ where $\lambda_i$ are Lagrange coefficients. 

\paragraph{Breach isolation}
We proactively quarantine confidentiality breaches by enforcing a privacy budget for each compute node. For this to work, we assume contract \tees have unforgeable host identities (e.g., the linkable EPID public key in SGX provides one). Key-manager nodes maintain a counter $\kappa_\computenode$ for each compute node $\computenode$ to record the number of queries. The counter is reset along with epoch advancement. Key-manager nodes fulfill a query only if $\kappa_\computenode < \kappa$ for some system parameter $\kappa$. With this in place, no matter how many \tees a breached compute node spawns, it can at most obtain $\kappa$ keys. In practice, requests to a depleted honest compute node can be redirected to other nodes, resulting in only a modest overhead.

\subsection{Atomic Delivery}
\label{sec:atomic}

Recall that \tee execution yields two messages:
$m_1$, which delivers the output to the caller, and $m_2$,
which delivers the state update to the blockchain,
both via adversarial channels.
As discussed in \cref{sec:atomic-overview}, it is critical to enforce atomic delivery of the two messages, i.e.
both $m_1$ and $m_2$ are delivered or the system has become permanently unavailable. Now we specify a protocol for atomic delivery.

Assuming a secure communication channel between a \tee and the calling client $\ekdparty$ (which in practice can be constructed with remote attestation),
we realize atomic delivery of $m_1$ and $m_2$ (defined above)
via the following two-phase protocol:
To initiate atomic delivery, \tee obtains a fresh key $\key$ from the key manager and sends an attested $m_1^c=\enc(\key, m_1)$ to $\ekdparty$ over a secure channel.
Once $\ekdparty$ acknowledges receipt of $m_1^c$, the \tee sends $m_2$ to the blockchain.
Finally, after seeing $\pi_{m_2}$, a proof of publication for $m_2$, \tee sends $\key$ to $\ekdparty$.

The above protocol realizes atomic delivery.
On the one hand, as a \tee can ascertain
the delivery of $m_2$ by verifying $\pi_{m_2}$,
$\key$ is revealed \emph{only if} $m_2$ is delivered.
On the other hand,
\emph{if} $m_2$ has been delivered,
$\key$ will be released eventually because at least one \tee is available and the key management protocol ensures that the availability of $\key$.

\section{Protocol Details and Security Proof}
\label{sec:model}

In this section, we specify $\protoekiden$, the protocol realization of \systemname{}. It aims to realize a Universal Composability (UC)~\cite{ucCanetti} ideal functionality $\fekiden$ that we defer to \cref{sec:supp_formalism} for lack of space and encourage the reader to consult. Looking ahead, $\protoekiden$ UC-realizes $\fekiden$.

\subsection{Preliminary and Notation}

\paragraph{Attested Execution}
To formally model attested execution on trusted hardware,
we adopt the ideal functionality $\funcsgx$ defined in~\cite{pass:formal}.
Informally, 
a party first loads a program $\sgxprog$ into a \tee with an $\msginstall$ message.
On a $\msgresume$ call, the program is run on the given input,
generating an output $\ekdoutput$ along with an attestation $\ekdatt = \sgxsig.\sig(\sksgx, (\sgxprog, \ekdoutput))$,
a signature under a hardware key $\sksgx$.
The public key $\pksgx$ can be obtained from $\funcsgx.\text{getpk()}$.
See~\cite{pass:formal} for details.

In practice it's useful to allow a \tee to output data that is not included in attestation. We extend $\funcsgx$ slightly to allow this: if a \tee program $\sgxprog$ generates a pair of output $(\ekdoutput_1,\ekdoutput_2)$, the attestation only signs $\ekdoutput_1$, i.e. $\ekdatt = \sgxsig.\sig(\sksgx, (\sgxprog, \ekdoutput_1))$. A common pattern is to include a hash of $\ekdoutput_2$ in $\ekdoutput_1$, to allow parties to verify $\ekdatt$ and $\ekdoutput_2$ separately. Similar technique is used in~\cite{zhang2017rem}.

Following the notation in~\cite{kosba2016hawk,sgp}, we use contract wrappers (defined in \cref{fig:enclavewrapper}) to abstract away routine functionality such as 
state encryption, key management, etc. A contract $\sf c$ augmented with the wrapper is denoted $\protowrapper{\sf c}$.

\paragraph{Blockchain}
$\fstorage[\funclink]$ (given in \cref{sec:supp_formalism})
defines a general-purpose append-only ledger implemented by common blockchain protocols (formally defined in \Cref{fig:idealstorage} in the Appendix).
The parameter $\funclink$ is a function that specifies the criteria for a new item to be added to the storage,
modeling the notion of transaction validity.
We retain the append-only property of blockchains but abstract away the inclusion of state updates in blocks.
We assume overlay semantics that associate blockchain data with $\ekdid$'s.
In addition to read and write interfaces,
$\fstorage$ provides a convenient interface by which clients can ascertain whether an item is included in the blockchain. In practice, this interface avoids the overhead of downloading the entire blockchain.

\paragraph{Parameterizing $\fstorage$}
\label{sec:ekidenstorage}

In \systemname{}, the contents of storage are parsed as an ordered array of \emph{state transitions},
defined as 
$\ekdstatetrans_i = (\hash(\ekdstate_{i-1}), \ekdstate_i, \sigma_i)$,
a tuple of a hash of the previous state, a new state, and a proof from \tee attesting to the correctness of a state transition.
(Note that as a performance optimization, large user input---e.g.\ training data in an ML contract--- may not be stored on chain.)
Storage can be interpreted as a special initial state followed by a sequence of state transitions: $\ekdstorage = ((\ekdprog,\ekdstate_0,\sigma_0), \set{\ekdstatetrans_i}_{i\ge 1}).$

For a state transition to be \emph{valid}, it must extends the latest state and the attestation must verify.
Formally, this is achieved by parameterizing $\fstorage$ with
a successor function $\funclink(\cdot,\cdot)$ such that
$\funclink(\ekdstorage, (h,\ekdstatenew,\ekdatt))=\true$
if and only if $h=\hash(\ekdstateprev)$ where $\ekdstateprev$ is the latest state in $\ekdstorage$ and $\sgxsig.\verify(\pksgx, \ekdatt, (h,\ekdstatenew))$.
This guarantees that at any time there is a single sequence of state transitions consistent with the view of each party, i.e. the chain of state transitions is fork-free.

\subsection{Formal Specification of the Protocol}
\label{sec:protocoldetails}

The \systemname{} protocol is formally specified in $\protoekiden$ (\cref{fig:protocolekiden}).
$\protoekiden$ relies on $\funcsgx$ and $\fstorage$, ideal functionality for attested execution and the blockchain.
$\protoekiden$ also use a digital signature scheme $\Sigma(\kgen, \sig, \verify)$, a symmetric encryption scheme $\sesys(\kgen,\enc,\dec)$ and an asymmetric encryption scheme $\aesys(\kgen,\enc,\dec)$.

\paragraph{Sharing state keys}
Each contract is associated with a set of keys. As discussed in \Cref{sec:keymanagement}, contract \tees delegate key management to key manager \tees. In $\protoekiden$, communication with key managers is abstracted away with the keyManager function.

\paragraph{Contract creation}
To create a contract in \systemname, a client $\ekdpartyi$ calls the \texttt{create} subroutine of a compute node \computenode with input $\ekdprog$, a piece of contract code.
\computenode loads the $\protowrapper{\ekdprog}$ into a \tee and starts the initialization by invoking the $\msgcreate$ call. As specified in \cref{fig:enclavewrapper}, the contract \tee creates a fresh contract $\ekdcid$, obtains fresh $(\pubkeyinput, \seckeyinput)$ pair and $\seckeycont$ from the key manager and generates an encrypted initial state $\ekdstate_0$ and an attestation $\ekdatt$. The attestation proves the $\ekdstate_0$ is correctly initialized and that $\pubkeyinput$ is the corresponding public key for contract $\ekdcid$. The compute node \computenode sends $(\ekdprog, \ekdcid, \ekdstate_0, \pubkeyinput, \ekdatt)$ to $\fstorage$ and waits for an receipt. $\computenode$ returns the contract $\ekdcid$ to $\ekdpartyi$, who will verify that contract $\ekdcid$ is properly stored on $\fstorage$.

\paragraph{Request execution}
To execute a request to contract $\ekdcid$, a client $\ekdpartyi$ first obtains the input encryption key $\pubkeyinput$ from $\fstorage$. Then $\ekdpartyi$ calls the \texttt{request} subroutine of \computenode with input $(\ekdcid, \ekdinputct)$, where $\ekdinputct$ is $\ekdpartyi$'s input encrypted with $\pubkeyinput$ and authenticated with $\spk_i$. \computenode fetches the encrypted previous state $\ekdstatect$ from $\fstorage$ and launches an contract \tee with code $\protowrapper{\ekdprog}$ and input $(\ekdcid, \ekdinputct, \ekdstatect)$.

As specified in \cref{fig:enclavewrapper}, if $\ekdpartysig$ verifies, the contract \tee decrypts $\ekdstatect$ and $\ekdinputct$ with keys obtained from the key manager and executes the contract program $\ekdprog$ to get $(\ekdstatenew, \ekdoutput)$. To ensure the new state and the output are delivered atomically, \computenode and $\ekdpartyi$ conduct an atomic delivery protocol as specified in \cref{sec:atomic}:
\compactlist{
    \item First the contract \tee computes $\ekdoutputct = \enc(\outputkey, \ekdoutput)$ and $\ekdstatect'=\ekdstatenewct$, and send both and proper attestation to $\ekdpartyi$ in a secure channel established by $\epk_i$.
    \item $\ekdpartyi$ acknowledges the reception by calling the \texttt{claim-output} subroutine of \computenode, which triggers the contract \tee to send $m_1=(\ekdstatect', \ekdoutputct, \sigma)$ to $\fstorage$. $\sigma$ protects the integrity of $m_1$ and cryptographically binds the new state and output to a previous state and a input, thus a malicious \computenode cannot tamper with it.
    \item Once $m_1$ is accepted by $\fstorage$, the contract \tee sends the decryption of $\ekdoutputct$ to $\ekdpartyi$ in a secure channel.
}

\subsection{Security of \protoekiden}

\Cref{thm:main} characterizes the security of $\protoekiden$. A proof sketch is given in \onlyinfullversion{sec:proof1}.

\begin{theorem}[Security of $\protoekiden$]
\label{thm:main}
Assume that $\funcsgx$'s attestation scheme $\Sigma_\text{TEE}$ and the digital signature $\Sigma$ are existentially unforgeable under chosen message attacks (EU-CMA),
that $\hash$ is second pre-image resistant, and that $\aesys$ and $\sesys$ are IND-CPA secure.
Then $\protoekiden$ securely realizes $\fekiden$ in the $(\funcsgx,\fstorage)$-hybrid model,
for static adversaries.
\end{theorem}

\subsection{Mitigating app-level leakage}
\label{subsub:application-levelleakage}
While \systemname{} protects within-\tee data, it is not designed to protect data at contract interfaces, i.e., data leakage resulting from the contract design.  (E.g., a secret prediction model may be ``extracted'' via client  queries~\cite{tramer2016stealing}.) Common approaches to minimizing such leakage, e.g., restricting requests based on requester identity and/or a differential-privacy budget~\cite{dwork2008differential, DBLP:journals/corr/JohnsonNS17}, require persistent counters. The monotonic counters in SGX are untrustworthy, however~\cite{DBLP:conf/uss/MateticAKDSGJC17}.

\systemname{} instead supports stateful approaches to mitigate application-level privacy leakage by enabling persistent application state---e.g., counters, total consumed differential privacy budget, etc.---to be maintained securely on chain. Moreover, the aforementioned atomic delivery guarantee ensures that the output is only revealed if this state is correctly updated.

\subsection{Performance Optimizations}

Given an additional mechanism for revocation, a simple modification {\em eliminates reliance on the IAS apart from initialization}. When initialized, an enclave creates a signing key $(\pk, \sk)$, and outputs $\pk$ with an attestation. Subsequently, attestations are replaced with signatures under $\sk$. Since $\pk$ is bound to the \tee code (by the initial attestation), signatures under $\sk$ prove the integrity of output, just as attestations do.
As with other keys, $(\pk, \sk)$ are managed by the key manager (c.f. \cref{sec:keymanagement}).

In \onlyinfullversion{sec:fullprotocol} we discuss an extended version of the protocol with several other performance optimizations.
\section{Implementation}
\label{sec:implementation}

\begin{figure*}
\protocolsidebyside
{$\protoekiden(\lambda, \aesys, \sesys, \Sigma, \set{\ekdpartyi}_{i\in[N]})$}
{
\underline{Clients $\ekdpartyi$:} \\
Initialize: $(\ssk_i, \spk_i) \sample \Sigma.\kgen(\secparam)$ \\
\t $(\esk_i, \epk_i) \sample \aesys.\kgen(\secparam)$ \\[1mm]
\onrecv $(\msgcreate, \ekdprog)$ from environment $\ekdenv$: \\
    \t $\ekdcid:=\text{create}(\ekdprog)$; assert $\ekdcid$ initialized on $\fstorage$ \\
    \t output $(\msgreceipt, \ekdcid)$ \\[1mm]
\onrecv $(\msgquery, \ekdcid, \ekdinput, \enclaveid)$ from environment $\ekdenv$: \\
    \t $\ekdpartysig := \sig(\ssk_i, (\ekdcid, \ekdinput))$ \\
    \t get $\pubkeyinput$ from $\fstorage$; \\
    \t let $\ekdinputct := \aesys.\enc(\pubkeyinput, (\ekdinput, \ekdpartysig))$ \\
    \t $(\ekdstatect', \ekdoutputct, \sigma):=\text{request}(\ekdcid, \ekdinputct)$ \\
    \t parse $\sigma$ as $(\ekdatt, \ekdinputhash, \ekdstateprevhash,\ekdoutputhash,\spk_i)$ \\
    \t assert $\hash(\ekdinputct) = \ekdinputhash$; assert $\ekdoutputct$ is correct by verifying $\sigma$ \\
    \t $o := \text{claim-output}(\ekdcid, \ekdstatect', \ekdoutputct, \sigma, \epk_i)$ \\
    \t \pccomment{retry if the previous state has been used by a parallel query} \\
    \t \pcif $o = \bot$ $\pcthen$ jump to the beginning of the $\msgquery$ call \\
    \t parse $o$ as $(\ekdoutputct', \ekdatt)$ \\
    \t assert $\sgxsig.\verify(\pksgx, \ekdatt, \ekdoutputct')$ \pccomment{$\pksgx := \funcsgx.\text{getpk()}$} \\
    \t output $\aesys.\dec(\esk_i, \ekdoutputct')$ \\[1mm]
\onrecv $(\msgread, \ekdcid)$ from environment $\ekdenv$: \\
    \t send $(\msgread, \ekdcid)$ to $\fstorage$ and relay output
}
{
\underline{Compute Nodes Subroutines (called by clients $\ekdpartyi$):} \\
\oninput create($\ekdprog$): \\
    \t send $(\msginstall, \protowrapper{\ekdprog})$ to $\funcsgx$, wait for $\enclaveid$ \\
    \t send $(\enclaveid, \msgresume, (\msgcreate))$ to $\funcsgx$ \\
    \t wait for $((\ekdprog, \ekdcid, \ekdstate_0, \pubkeyinput), \ekdatt)$ \\
    \t send $(\msgwrite, (\ekdprog, \ekdcid, \ekdstate_0, \pubkeyinput, \ekdatt))$ to $\fstorage$ \\
    \t wait to receive $(\msgreceipt, \ekdcid)$ \\[1mm]
\oninput request($\ekdcid, \ekdinputct$): \\
    \t send $(\msgread, \ekdcid)$ to $\fstorage$ and wait for $\ekdstatect$ \\
    \t \pccomment{non-existing $\enclaveid$ is assumed to be created transparently} \\
    \t send $(\enclaveid, \msgresume, (\msgquery, \ekdcid, \ekdinputct, \ekdstatect))$ to $\funcsgx$ \\
    \t receive $((\msginterimoutput, \ekdinputhash, \ekdstateprevhash,\ekdstatect',\ekdoutputhash,\spk_i), \ekdatt, \ekdoutputct)$ \\
    \t \pccomment{$\ekdatt = \sgxsig.\sig(\sksgx, (\ekdinputhash, \ekdstateprevhash,\ekdstatect',\ekdoutputhash,\spk_i))$} \\
    \t let $\sigma := (\ekdatt, \ekdinputhash, \ekdstateprevhash,\ekdoutputhash,\spk_i)$ \\
    \t \pcreturn $(\ekdstatect', \ekdoutputct, \sigma)$ \\[1mm]
\oninput claim-output$(\ekdcid, \ekdstatect', \ekdoutputct, \sigma, \epk_i)$: \\
    \t send $(\msgwrite, \ekdcid, (\ekdstatect', \sigma))$ to $\fstorage$ \\
    \t \pcif receive $(\msgreject, \ekdcid)$ from \fstorage $\pcthen$: return $\bot$ \\
    \t send $(\enclaveid, \msgresume, (\msgclaimoutput, \ekdstatect', \ekdoutputct, \sigma, \epk_i))$ to $\funcsgx$ \\
    \t receive $(\msgfinaloutput,\ekdoutputct', \ekdatt)$ from $\funcsgx$ or abort \\
    \t \pcreturn $(\ekdoutputct', \ekdatt)$
}
\caption{\systemname Protocol. The contract \tee program $\protowrapper{\ekdprog}$ is defined in \Cref{fig:enclavewrapper}, in \Cref{sec:supp_formalism}.}
\label{fig:protocolekiden}
\end{figure*} 
We implemented an \systemname{} prototype in about $7.5$k lines of Rust.
We also implemented a compiler that automatically builds
contracts into executables that can be loaded into a compute node,
using the Rust SGX SDK~\cite{Ding:2017:PRS:3133956.3138824}.

\systemname{} is compatible with many existing blockchains.
We have built one end-to-end instantiation, \emph{\systemimpl{}},
with a blockchain extending from Tendermint~\cite{kwon2014tendermint}, which required no changes to Tendermint.

\subsection{Programming Model}
\label{sec:impl:programming}

We support a general-purpose programming model for specifying contracts.
A contract registers a mutable struct as its state, which \systemname{}
transparently serializes, encrypts, and synchronizes with the blockchain after method calls.
Contract methods must be deterministic and terminate in bounded time.
Within this model, we implemented two smart-contract programming environments.
In the Rust backend, developers can write contracts using a subset of the 
Rust programming language, and thus benefit from a range of open source libraries.
We also ported the Ethereum Virtual Machine (EVM), thereby supporting
any contract written for the Ethereum platform.
The system currently does not support calling contract functions from another contract. We leave this for future work.

\subsection{Applications}
\label{subsec:apps}

\begin{figure*}[t]
\centering
\begin{tabular}{l|l|r|l|l}
\toprule
Application         & Language & LoC  & \textbf{Secret Input/Output}   & \textbf{Secret State} \\
\toprule
Machine Learning    & Rust     & $806$  & Training data, predictions     & Model             \\
Thermal Modeling    & Rust     & $621$  & Sensor data, temperature       & Building model    \\
Token               & Rust     & $514$  & Transfer (from, to, amount)     & Account balances  \\
Poker               & Rust     & $883$  & Players' cards                 & Shuffled deck     \\
Ethereum VM         & Rust     & $1411$ & Input and output               & Contract state    \\
\midrule
CryptoKitties       & EVM Bytecode & $54^*$   & Random mutations               & Breeding algorithm \\
Origin Demo         & Solidity, JS & $19^*$   & Purchase orders                & Purchase history   \\
\bottomrule
\end{tabular}
\caption{
    \systemname{} smart contracts.
    For each, we specify the implementation language,
    development effort (LoC),
    as well as secret inputs, outputs, and state.
    Secret inputs and outputs are only accessible to the contract and the invoking user.
    Secret state is only accessible to the contract.
    For the EVM, we only include the cost of porting Parity-Ethereum's runtime.
    For CryptoKitties and Origin Demo, we only include LoC specific to porting, as marked by $*$.
}
\label{fig:apps}
\end{figure*}

We now describe several different applications we developed to show the versatility
of \systemname{}'s programming model.
Figure~\ref{fig:apps} highlights the secret state and application complexity of each contract.

\paragraph{Machine Learning Contracts}

To demonstrate shared learning on secret data, we implemented two example contracts:
(i) credit scoring based on financial records~\cite{baesens2003benchmarking} and
(ii) predicting the likelihood of heart disease based on medical records~\cite{sajda2006machine}.
In both of these, we used a version of the rusty-machine~\cite{rusty-machine} machine learning library, which we ported to run inside our contracts.
The training data given to these example contracts is treated as sensitive data (we use data from the UCI machine learning repository~\cite{Lichman:2013} in our experiments) and never exposed as plaintext outside the contract.

Our example contracts train the models with added noise for differential privacy.
This prevents information about the training data from leaking~\cite{shokri2017membership} during inference.
Ekiden's private computation guarantee allows the noise to be added centrally, which results in better accuracy and utility at the same level of privacy, compared to having clients add noise before submitting their data~\cite{dwork2014algorithmic}.
Additionally, after training, multiple compute nodes can run serve inference requests at high capacity without affecting correctness or privacy.

\paragraph{Smart Building Thermal Modeling}
We ported an implementation of non-linear least squares, which is used to predict temperatures
based on time series thermal data from smart buildings~\cite{dewson1993least}.
We have deployed this smart contract to train a shared model across real-time data from select buildings in Berkeley, CA.
These buildings sample their temperature sensors every 20 seconds, generating data used to update the predictive model.
\systemname{} allows the contract to run its model while keeping the sensor data and model secret, demonstrating that our system is sufficiently responsive
for highly interactive workloads in an online setting.

\paragraph{Tokens}

The most popular kind of Ethereum contract is the ERC20 token standard. Using the Ethereum port (\Cref{sec:impl:programming}), we can run existing ERC20 token contracts. We also implemented a token contract written directly in Rust, which yields moderate performance improvement (see~\Cref{sec:evaluation}). In either case, \systemname{} automatically provides privacy and anonymity, which the contract would not receive on the Ethereum mainnet.
The secret state in the token the account balance for each user.

\paragraph{Poker}
We also implemented a poker contract, where users take turns submitting their actions to the contract, and the smart contract contains all of the game logic for shuffling and (selectively) revealing cards.
 Poker is a common benchmark application for blockchain systems and secure multi-party computation called \textit{mental poker}~\cite{bentov2017instantaneous,kumaresan2015use,kumaresan2016amortizing,andrychowicz2014secure}. 
Ekiden is significantly more robust than these prior implementations in how it handles player aborts. 
In most mental poker, if a party aborts, its secret hand cannot be reconstructed by others, so the game aborts.
Handling faults in secure multi-party computation requires application-specific changes to the cryptographic protocol~\cite{castella2005dropout}.
Because \systemname persists state to the blockchain after each action, and can be accessed from any enclave, secret cards can still be revealed if a player aborts.

\paragraph{CryptoKitties}

CryptoKitties~\cite{cryptokitties} is an Ethereum game that allows users to breed virtual cats, 
which are stored on chain as ERC721 tokens~\cite{erc721}.
Each cat has a unique set of genes that determine its appearance and therefore its value.
The traits of offspring are determined by a smart contract that mixes the genes of its parents.
The source code of the gene mixing contract is not publicly available: The game developers aimed to make the breeding process unpredictable.

We obtained the bytecode for the gene mixing contract from the Ethereum blockchain and executed it using our \systemname{} EVM port. 
We verified correct behavior by reproducing real transactions from the Ethereum network.
This example demonstrates that \systemname{} can execute an Ethereum contract even when source code is not available.
Further, \systemname{} can provide unique benefits for smart contracts requiring secrecy or unpredictability such as CryptoKitties. These properties are difficult to achieve with Ethereum. E.g., the CryptoKitties gene mixing algorithm has been reverse-engineered~\cite{erays}, which allows strategic players to optimize their chance of breeding cats with rare traits, thus undermining the game's ecosystem. By contrast, an \systemname{} contract has access to a source of randomness in hardware and allows secret elements of a game's algorithm to be stored in encrypted state.

\paragraph{Origin}
Origin~\cite{originprotocol} is a platform for building online marketplaces on top of Ethereum. We ported a demo application which allows users to list and purchase items with Ether.
This application further demonstrates that development frameworks built for Ethereum can be easily used by \systemname{}: the smart contracts used in the demo work without modification; we were able to integrate the rest of the demo, namely, a user-facing web server, with minor modifications.
Built on \systemname{}, users' transaction history in the blockchain are kept private, and transactions are confirmed faster than on Ethereum.
\section{Evaluation}
\label{sec:evaluation}

In this section, we present evaluation results for  end-to-end latency and
peak throughput. We evaluated the five applications of \Cref{subsec:apps}:
a Rust token contract \textbf{Token}, implementing an ERC20-like token in the Rust language,
two Ethereum contracts, \textbf{ERC20} and \textbf{CryptoKitties}, running in the ported EVM, and two machine learning applications, \textbf{Credit} and \textbf{Thermal}.
Compared to an ERC20 contract on Ethereum mainnet,
\systemimpl{} can support a token contract
with 600x greater throughput,
400x less latency, at 1000x less monetary cost.
While we expect some mild performance degradation when deployed with a larger scale blockchain,
our performance optimizations significantly reduce the effect of the blockchain's speed, as shown below.
Furthermore, we demonstrate that \systemname{} can efficiently support computation-intensive workloads such as machine learning applications
which would be cost-prohibitive on Ethereum.
We also quantify the performance gains from each of the optimizations described in \onlyinfullversion{sec:fullprotocol}.
We show that batching, caching, and a write-ahead log improve
performance and reduce the network costs of synchronizing state with the blockchain.

\subsection{Experimental Setup}
To evaluate the performance of \systemimpl{},
we ran experiments with four consensus nodes hosted on Amazon EC2 across different availability zones
and one compute node (with a Core i7-6500U CPU with 8GB of memory) hosted locally,
as EC2 does not offer SGX-enabled instances at the time of writing.
Transactions are only run once on the compute node ($K=1$).
Each consensus node was run on an \texttt{t2.medium} instance, with 2 CPU cores and 4 GB of memory.
As shown in \Cref{sec:eval:throughput}, we do not expect throughput performance to be significantly impacted
by a larger slower blockchain, because many transactions can be compressed into a single write onto the blockchain.
By separating execution from consensus, these layers can work in parallel.
However achieving consensus among a larger group of consensus nodes will result in higher end-to-end latencies.

\subsection{End-to-End Latency}
\begin{figure}
\begin{center}
    \includegraphics[width=0.45\textwidth]{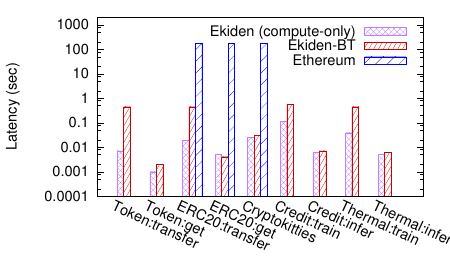}
\end{center}
\vspace{-1.75em}
\caption{
    End-to-end latency of client requests for various contracts, plotted on a log scale.
    Running Rust token and ERC20 token contracts on \systemimpl{} yields transactions
    2-5 orders of magnitude faster than Ethereum.
    Read-write transactions on the \systemimpl{} blockchain take about a second, dominated by the underlying blockchain.
    Caching avoids writes to the blockchain for read-only transactions (e.g. \texttt{get}).
    We only compare Ethereum for the ERC20 contract, as there are no comparable machine learning contracts on Ethereum.
}
\label{fig:latency}
\end{figure}

\Cref{fig:latency} shows end-to-end latency for calling the token, CryptoKitties,
and machine learning contracts, plotted on a log scale.
For the ``\systemimpl{}'' plot, we start our timer when the client triggers a request
and end when the smart-contract response, committed on chain, is decrypted.
For read-only transactions like ``Token:get'' or ``Credit:infer'', compute nodes use a locally cached copy of state.
Writes to the \systemimpl{} blockchain take up to a second to confirm.
Latencies in \systemname{} are dominated by the time to commit on chain.
This relative cost is lower for compute-intensive workloads like machine learning training.
For comparison, we include a bar (``compute-only'') that measures computation time only. 

For the three transactions that could be run on the Ethereum network,
we plot the publicly reported block rates of the Ethereum mainnet in March 2018~\cite{etherscan},
which represents the optimistic case that transactions are incorporated in the next block.
Compared to the proof-of-work protocol used in Ethereum,
\systemimpl{} has 2-3 orders of magnitude faster confirmations,
in part due to the use of a faster blockchain.
For the ERC20 token, which runs on the EVM in \systemimpl{},
we see similar performance to the Rust token contract,
because both use the same consensus protocol.

\subsection{Throughput}
\label{sec:eval:throughput}
\begin{figure}
\begin{center}
    \includegraphics[width=0.45\textwidth]{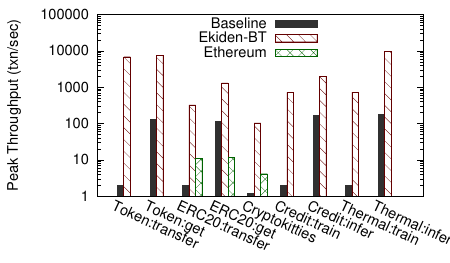}
\end{center}
\vspace{-1.75em}
\caption{
    Throughput comparison across contracts and systems.
   Our baseline reads and writes to a blockchain for every request. Throughput is limited by blockchain performance.
    Our optimizations improve performance by 2--4 orders of magnitude over the baseline,
    with more advantage for read-write operations on contracts with large state (e.g. Token).
    In-EVM operations incur about 10x higher cost compared to our Rust token. For ERC20,
    we achieve 1--2 orders of magnitude higher performance than Ethereum.
}
\label{fig:throughput}
\end{figure}

To measure \systemimpl{}'s peak performance,
we conducted an experiment with 1000 clients,
each sending 100 serialized requests to a compute node.
For each data point, we disregard the first and last 10\% of requests,
averaging the stable performance under stress.
\Cref{fig:throughput} shows the results
for the token, CryptoKitties, and machine learning contracts.
For the baseline, we implement the simplest \systemimpl{} protocol,
where each request triggers a full state checkpoint on our blockchain.
In the ``\systemimpl{}'' bar, we include our optimizations,
as described in \onlyinfullversion{sec:fullprotocol}.
Batching compresses multiple state checkpoints into a single commit on the blockchain.
We then cache the latest state on compute nodes and use a write-ahead log for state updates.
Our optimizations have the greatest benefit for read-write operations, like \texttt{transfer}.
They have less benefit for contracts with smaller states, such as the machine learning contract
with small models.
Conversely, writes to the blockchain significantly impact performance for read-write transactions,
compared to read-only transactions with cached state.
For comparison on the transactions that could be run on the Ethereum network,
we plot the publicly reported transaction throughput of the Ethereum mainnet in March 2018~\cite{etherscan}.
Because CryptoKitties incurs higher computational cost, we can fit fewer transactions in a block due to the gas limit, compared to ERC20 transactions.

\subsection{Impact of Consensus on Throughput}
\begin{figure}
\begin{center}
    \includegraphics[width=0.45\textwidth, height=0.23\textwidth]{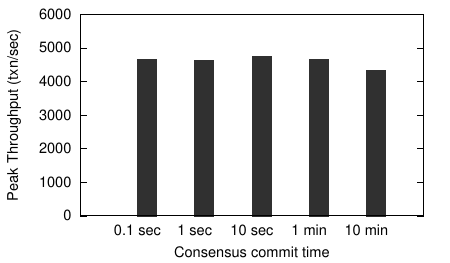}
\end{center}
\vspace{-1.00em}
\caption{
   Peak throughput performance of token transfers
   under different consensus layer commit times.
   Because contract execution occurs in parallel to state agreement,
   we show that good throughput performance for a wide range of commit times on the consensus layer.
   We expect \systemname{} to perform well on a variety of blockchains.
}
\label{fig:varyconsensus}
\end{figure}

To understand the impact of using different consensus protocols
with \systemname{}, we measured peak throughput performance of token transfers
as a function of the time to commit state to the blockchain.
In order to simulate slower consensus protocols, we inject a variable
delay for writes to the consensus nodes.
Figure~\ref{fig:varyconsensus} shows that token transfers have good performance for a wide range of commit latencies seen in popular blockchains.

Because state is cached at compute nodes,
compute nodes can opportunistically execute new transactions
without waiting for a response from consensus nodes.
Periodically, compute nodes asynchronously commit the state to
the blockchain, as defined by the batch size.
By separating contract execution from agreement on state,
the layers can operate in parallel.

In contrast,
Ethereum transactions are broadcast to all miners. Miners execute transactions sequentially, and all contracts are serialized onto a single blockchain.
At the time of writing, there are 36974 ERC20 token contracts, all using the Ethereum blockchain~\cite{etherscan}. In contrast, \systemname{} parallelizes contracts across compute nodes, eliminating computational bottlenecks for better performance.
However, implementation of full cross-contract calls remains future work.

\subsection{Transaction Costs}
In March 2018 on Ethereum, it cost 52K gas (\$0.17 USD) to perform a transfer on an ERC20 token contract
and 130K gas (\$0.39 USD) to compute the breeding algorithm on CryptoKitties~\cite{ethgasstation}.
By contrast, IBM rents machines with Intel SGX processors useable by \systemname{} for \$260.00 per month. These can do a token transfer in 2ms and CryptoKitties breeding in 100ms,
at a cost of roughly $10^{-7}$ and $10^{-5}$ dollars respectively, and a cost of $10^{-5}$ dollars for each call to \textit{train} in our machine learning contract.
For these contracts, the cost to commit state to the Ethereum blockchain ranges from
\$0.0688 for CryptoKitties to \$1.92 to store a 1KB machine learning model.
Because \systemname{} can compress results from multiple requests into a single write to the blockchain,
our system has a total cost vastly less than that of on-chain execution.
There are no current public deployments of Tendermint for comparison.

\section{Related Work}
\label{sec:related}

\noindent {\bf Confidential smart contracts:}
Hawk~\cite{kosba2016hawk} is a smart contract system that provides confidentiality by
executing contracts off-chain and posting only zero-knowledge proofs on-chain.
As the zero-knowledge proofs in Hawk (zk-SNARKs) incur very high computational overhead, \systemname{} is significantly faster. Additionally, Hawk was designed for a single compute node (called the ``manager''), and thus cannot (as designed) offer high availability.
While \systemname{} does require trust in the security of Intel SGX, Hawk's ``manager'' must be trusted for privacy.
Hawk supports only a limited range of contract types, not the general functionality of \systemname{}. 

The idea of combining ledgers with trusted hardware for smart contract execution is briefly mentioned in Hawk and also treated in~\cite{fairnessChoudhuri, kaptchuk2017giving}.
~\cite{fairnessChoudhuri} combines blockchain with \tee to achieve one-time programs that resemble  smart contracts but only aim for a restricted functionality (one-shot MPC with $N$ parties providing input).~\cite{kaptchuk2017giving} includes a basic prototype, but omits critical system design issues; e.g., its permissionless ``proof-of-publication'' overlooks the technical difficulties arising from lack of trusted wall-clock time in enclaves. 

\systemname{} is also closely related to and influenced by Hyperledger Private Data Objects (PDO)~\cite{bowman2018private} from Intel. PDOs use smart contracts, executed in SGX enclaves, to mediate access to data objects shared amongst mutually distrusting parties. 
To the best of our knowledge, PDOs target permissioned and managed settings (requiring, e.g., special-purpose validation rules), while \systemname{} supports permisionless and open settings as well. This leads to key technical differences. For example, PDO uses a set of Provisioning Services to store encryption keys without worrying about availability risk, which cannot be easily realized in the \systemname{} setting where churn is possible. In contrast, \systemname{} uses a secret-sharing-based key management protocol that tolerates churn and allows flexible committee reconfiguration. 

The Microsoft Coco Framework~\cite{ms-coco} is concurrent and independent work to port existing smart contract systems,
such as Ethereum, into an SGX enclave.
To the best of our knowledge, only a whitepaper containing a high-level overview has been produced.
No details of a protocol or implementation have yet been released.

\vspace{0.05in}
\noindent {\bf Blockchain transaction privacy:}
\systemname{}'s goals relate to mechanisms for enhancing transaction privacy on public blockchains.
Maxwell proposed a confidential transaction scheme~\cite{maxwellct} for Bitcoin that conceals transaction amounts,  but not  identities.
Zerocash~\cite{DBLP:conf/sp/Ben-SassonCG0MTV14} as well as Cryptonote~\cite{sun2017ringct,van2013cryptonote},
Solidus~\cite{DBLP:conf/ccs/CecchettiZJKJS17},
and Zerocoin~\cite{DBLP:conf/sp/MiersG0R13} provides stronger confidentiality guarantees by concealing identities. 
These schemes, however, 
do not support smart contracts.

\vspace{0.05in}
\noindent {\bf Privacy-preserving systems based on trusted hardware:}
Trusted hardware, particularly Intel SGX,
has seen a wide spectrum of applications in distributed systems. M$^2$R~\cite{dinh2015m2r}, VC3~\cite{schuster2015vc3}, Opaque~\cite{zheng2017opaque}
and Ohrimenko \etal~\cite{ohrimenko2016oblivious}
leverage SGX to offer privacy-preserving data analytics and machine learning with various security guarantees,
Ryoan~\cite{hunt2016ryoan} is a distributed sandbox platform
using SGX to confine privacy leakage from untrusted applications that process sensitive data.
These systems do not address state integrity and confidentiality over a long-lived system.
In comparison, \systemname{} provides a stronger integrity and availability guarantees by persisting
contract states on a blockchain.

\vspace{0.05in}
\noindent {\bf Blockchains for verifiable computations and secure multi-party computations:}
Several related works offer blockchain-based guarantees of computation integrity, but cannot guarantee privacy~\cite{luu2015demystifying,teutsch2017scalable,teutsch2017interactive}.
Other works have used a blockchain for fairness in MPC by requiring parties to forfeit security deposits if they abort~\cite{bentov2017instantaneous,kumaresan2015use,kumaresan2016amortizing,andrychowicz2014secure,zyskind2015decentralizing, fairnessChoudhuri}. Compared to these, \systemname{} can guarantee that all data can be recovered if \emph{any} compute node remains online. TEE-based computation is also far more performant than MPC. A theoretical scheme~\cite{goyal2017overcoming} combines witness encryption with proof-of-stake blockchains to achieve one-time programs that resemble smart contracts but avoid use of trusted hardware. This scheme is regrettably even more impractical than MPC.

\section{Conclusion}
\label{sec:conclusion}

\systemname{} demonstrates that blockchains and trusted enclaves have complementary security properties that can be combined effectively to provide a powerful, generic platform for \smartcontracts. The result is a compelling programming model that overcomes significant challenges in blockchain smart contracts. We show that \systemname{} can be used to implement a variety of secure decentralized applications that compute on sensitive data.

In future work we plan to extend \systemname{} to operate under a stronger threat model, leveraging techniques such as
secure multi-party computation~\cite{lindell2009secure,fairnessChoudhuri,andrychowicz2014secure},
to protect the system's more critical features, such as key management and coordination across compute nodes.
Coordination can also facilitate parallelism in contract execution,
merging concurrent output from  multiple enclaves to obtain still higher performance from \systemname{}.

\section*{Acknowledgments}
We wish to thank Intel, and Mic Bowman in particular, for ongoing research discussions and generous support of a number of aspects of this work. Our discussions regarding Intel's PDO system illuminated important technical challenges in \systemname{} and influenced and helped us refine its design.

We also wish to thank Iddo Bentov, Joe Near, Chang Liu, Jian Liu, and Lun Wang for their helpful feedback and discussion. We also thank Pranav Gaddamadugu and Andy Wang for their contributions to application development. This material is in part based upon work supported by the  Center for Long-Term Cybersecurity, DARPA (award number N66001-15-C-4066) IC3 industry partners, and the National Science Foundation (NSF award numbers TWC-1518899 CNS-1330599, CNS-1514163, CNS-1564102, CNS-1704615, and ARO W911NF-16-1-0145). This work was also supported in part by FORCES (Foundations Of Resilient CybEr-Physical Systems), which receives support from the National Science Foundation (NSF award numbers CNS-1238959, CNS-1238962, CNS-1239054, CNS-1239166). Any opinions, findings, and conclusions or recommendations expressed in this material are those of the author(s) and do not necessarily reflect the views of the National Science Foundation.
 
{
\footnotesize
\bibliographystyle{IEEEtranS}
\bibliography{biblio.bib}
}

\appendix

\iffullpaper
\subsection{Supplementary Formalism}
\label{sec:supp_formalism}

\subsubsection{Ideal Blockchain}
\label{subsec:fblockchain}

We specify the ideal functionality for a blockchain in \cref{fig:idealstorage}.

\begin{figure}
\centering
\protocol
{$\fstorage[\funclink]$}
{Parameter: successor relationship $\funclink:\bin^*\times\bin^*\to\bin$ \\
\onrecv $(\msginit)$: $\ekdstorage := \emptyset$ \\
\onrecv $(\msgread, \ekdid)$: output $\ekdstorage[\ekdid]$, or $\bot$ if not found \\
\onrecv $(\msgwrite, \ekdid, \ekdinput)$ from $\ekdparty$: \\
    \t let $\ekdvalue := \ekdstorage[\ekdid]$, set to $\bot$ if not found \\
    \t \pcif $\funclink(\ekdvalue, \ekdinput) = 1 \pcthen$ \\
    \t\t $\ekdstorage[\ekdid] := \ekdvalue\,\|\,(\ekdinput, \ekdparty)$; output $(\msgreceipt, \ekdid)$ \\
    \t \pcelse output $(\msgreject, \ekdid)$ \\
\onrecv $(\msgcheckreceipt, \ekdid, \ekdvalue)$: \\
\   \t \pcif $\ekdvalue \in \ekdstorage[\ekdid] \pcthen$ output $\true$ \pcelse output $\false$
}\caption{Ideal blockchain.
The parameter $\funclink$ defines the validity of new items.
A new item can only be appended to the storage if the evaluation of $\funclink$ outputs $1$.}
\label{fig:idealstorage}
\end{figure}

\subsubsection{Ideal functionality $\fekiden$}
\label{sec:fekiden}

\begin{figure}
\centering
\protocol
{$\fekiden(\secpar, \ell, \set{\ekdpartyi}_{i\in[N]})$}
{
Parameter: leakage function $\ell:\bin^* \to \bin^*$ \\
\onrecv ($\msginit$): $\ekdstorage := \emptyset$\\[1mm]
\pccomment{Create a new contract}\\
\onrecv $(\msgcreate, \ekdprog)$ from $\ekdpartyi$ for some $i\in[N]$: \\
    \t $\ekdcid \sample \bin^\secpar$ \\
    \t notify $\ekdadv$ of $(\msgcreate, \ekdpartyi, \ekdcid, \ekdprog)$; block until $\ekdadv$ replies\\
    \t $\ekdstorage[\ekdcid]:=(\ekdprog, \vec{0})$ \\
    \t send a public delayed output $(\msgreceipt,\ekdcid)$ to $\ekdpartyi$ \\[1mm]
\pccomment{Send queries to a contract} \\
\onrecv $(\msgquery, \ekdcid, \ekdinput, \enclaveid)$ from $\ekdpartyi$ for some $i\in[N]$: \\
    \t notify $\ekdadv$ of $(\msgquery, \ekdcid, \ekdpartyi, \ell(\ekdinput))$ \\
    \t $(\ekdprog, \ekdstate, \_) := \ekdstorage[\ekdcid]$; abort if not found \\
    \t $(\ekdoutput, \ekdstate') := \ekdprog(\ekdpartyi, \ekdinput, \ekdstate)$ \\
    \t let $\ell_{\ekdstate} = \ell(\ekdstate)$ \\
    \t notify $\ekdadv$ of $(\ekdcid, \ell_{\ekdstate'}, \ell(\ekdoutput), \enclaveid)$ \\
    \t wait for $\msgok$ from $\ekdadv$ and halt if other messages received \\
    \t update $\ekdstorage[\ekdcid] := (\ekdprog, \ekdstate', \ell_{\ekdstate'})$ \\
    \t send a secret delayed output $\ekdoutput$ to $\ekdpartyi$ \\[1mm]
\pccomment{Allow public access to encrypted state} \\
\onrecv $(\msgread, \ekdcid)$ from $\ekdpartyi$ for some $i\in[N]$: \\
    \t $(\_, \_, \ell_{\ekdstate}) := \ekdstorage[\ekdcid]$; abort if not found \\
    \t send $\ell_{\ekdstate}$ to $\ekdpartyi$ \\
    \t if $\ekdpartyi$ is corrupted: send $\ell_{\ekdstate}$ to $\ekdadv$
}\caption{The ideal functionality of \systemname.}
\label{fig:idealekiden}
\end{figure}

We specify the security goals of \systemname in the ideal functionality $\fekiden$ defined in \Cref{fig:idealekiden}.

$\fekiden$ allows parties to create contracts and interact with them.
Each party $\ekdpartyi$ is identified by a unique id simply denoted $\ekdpartyi$. Parties send messages over \emph{authenticated channels}. To capture the allowed information leakage from the encryption, we follow the convention of~\cite{ucCanetti} and parameterize $\fekiden$ with a leakage function $\ell(\cdot)$. We use the standard \emph{delayed output} terminology~\cite{ucCanetti} to model the power of the network adversary. Specifically, when $\fekiden$ sends a delayed output $\ekdoutput$ to $\ekdparty$, this means that $\ekdoutput$ is first sent to the adversary $\ekdadv$ and forwarded to $\ekdparty$ after acknowledgement by $\ekdadv$. If the message is secret, only the allowed amount of leakage (i.e., that specified by the leakage function) is revealed to $\sdv$.

A $\ekdprog$ is a user-provided program. Each smart contract is associated with a piece of persistent storage where the contract code and $\ekdstate$ can be stored. The storage is public; therefore $\fekiden$ allows any party, including $\ekdadv$, to read the storage content. The information leakage through such reading is also defined by the leakage function $\ell$.

Users can send queries to $\fekiden$ to execute the contract code
with user-provided input. The execution of a contract will result
in a secret output (denoted $\ekdoutput$) returned to the invoker and a secret transition to a new contract state (denoted $\ekdstate'$), equivalent intuitively to black-box contract execution (modulo leakage).
Although any party may send messages to the contract,
the contract code can enforce access control
based on the calling pseudonym passed to the contract.

\paragraph{Corruption model}
$\fekiden$ adopts the standard corruption model of~\cite{ucCanetti}.
$\ekdadv$ can corrupt any number of
clients, and up to all but one contract executors.
When $\ekdadv$ corrupts a \tee (or similarly a party),
$\ekdadv$ sends the message (``corrupt'', $\enclaveid$) to $\fekiden$.
If a query includes an invalid \tee id,
$\fekiden$ aborts if instructed by $\ekdadv$.
Otherwise the ideal functionality ignores $\enclaveid$s, which are included in $\fekiden$ only as a technical requirement to ensure interface compatibility with $\protoekiden$, given below.

\subsubsection{Contract \tee wrapper}

The contract TEE wrapper $\protowrapper{\ekdprog}$ is specified in \cref{fig:enclavewrapper}.

\begin{figure}[t]
\protocol{Contract \tee wrapper $\protowrapper{\ekdprog}$}{
\oninput $(\msgcreate):$ \\
    \t $\ekdcid := \hash(\ekdprog)$ \\
    \t $(\pubkeyinput, \seckeyinput):= \text{keyManager}(\msginputkey)$ \\
    \t $\seckeycont := \text{keyManager}(\msgstatekey)$ \\
    \t $\ekdstate_0 = \sesys.\enc(\seckeycont, \vec{0})$ \\
    \t $\pcreturn (\ekdprog, \ekdcid, \state_0, \pubkeyinput)$ \\[1mm]
\oninput $(\msgquery, \ekdcid, \ekdinputct, \ekdstatect)$: \\
    \t \pccomment{retrieve $\seckeyinput, \seckeycont$ from a key manager as above} \\
    \t ($\ekdinput, \ekdpartysig) := \aesys.\dec(\seckeyinput, \ekdinputct)$ \\
    \t assert $\verify(\ekdpartysig, \spk_i, (\ekdcid,\ekdinput))$ \pccomment{$\spk_i$ is publicly known} \\
    \t $\ekdstateprev := \sesys.\dec(\seckeycont, \ekdstatect)$ \\
    \t $\ekdstatenew, \ekdoutput := \ekdprog(\ekdstateprev, \ekdinput, \spk_i)$ \\
    \t $\ekdstatect':= \sesys.\enc(\seckeycont, \ekdstatenew)$ \\
    \t \pccomment{initiate atomic delivery} \\
    \t $\outputkey := \text{keyManager}(\msgoutputkey)$ \\
    \t $\ekdoutputct := \sesys.\enc(\outputkey, \ekdoutput)$ \\
    \t let $\ekdinputhash:=\hash(\ekdinputct)$, $\ekdstateprevhash:=\hash(\ekdstatect)$, $\ekdoutputhash=\hash(\ekdoutputct)$ \\
    \t $\pcreturn ((\msginterimoutput, \ekdinputhash, \ekdstateprevhash,\ekdstatect',\ekdoutputhash,\spk_i), \ekdoutputct)$ \\[1mm]
\oninput $(\msgclaimoutput, \ekdstatect', \ekdoutputct, \sigma, \epk_i)$: \\
    \t parse $\sigma$ as $(\ekdatt, \ekdinputhash, \ekdstateprevhash,\ekdoutputhash,\spk_i)$ \\
    \t assert $\hash(\ekdoutputct) = \ekdoutputhash$ \\
    \t send $(\msgcheckreceipt, \ekdcid, (\ekdstatect',\sigma))$ to $\fstorage$ \\
    \t receive $\true$ from $\fstorage$ or abort\\
    \t $\outputkey := \text{keyManager}(\msgoutputkey)$ \\
    \t $\ekdoutput := \sesys.\dec(\outputkey, \ekdoutputct)$ \\
    \t $\pcreturn (\msgfinaloutput,\aesys.\enc(\epk, \ekdoutput))$
}\caption{Contract \tee wrapper.}
\label{fig:enclavewrapper}
\end{figure} \subsection{Proof of Publication}

The protocol for proof of publication is specified in \cref{fig:proofofpublication}.

\begin{figure}
\protocol{Proof of Publication of $m$ between verifier $\enclaveparty$ and prover $\ekdparty$}
{
\underline{Parameters:} \\
$n_c$: publication of $m$ needs at least $n_c$ confirmation \\
$CB:$ a recent checkpoint block \\
$\delta(CB)$: difficulty of $CB$ \\
$\tau$: expected block interval of main chain \\
$\epsilon$: slackness factor\\[2mm]
\underline{\bf Verifier $\enclaveparty$ (a contract \tee):} \\
$t_1\gets$ \tee.timer() \\
$r \sample \bin^{\secpar}$ \\
send $(m,r)$ to $\ekdparty$ \\
receive $\pi_{(m,r)}=(CB, B_1, \cdots, B_n)$ from $\ekdparty$ \\
$t_2\gets$ \tee.timer() \\
\pcif $\pi_{(m,r)}$ is not a valid chain, output $\false$ \\
let $B_i \in \pi_{(m,r)}$ be the block that contains $(m,r)$, output $\false$ if $\nexists B_i$ \\
\pcif $B_i$ has less than $n_c$ confirmation, i.e. $n-i < n_c$, output $\false$ \\
\pcif any $B\in \pi_{(m,r)}$ has a lower difficulty than $\delta(CB)$, output $\false$ \\
\pcif $t_2 - t_1 < (n-i)\times \tau \times \epsilon$: output $\true$ and update checkpiont $CB=B_n$ \\
\pcelse: output $\false$ \\[2mm]
\underline{\bf Prover $\ekdparty$:} \\
\onrecv $(m,r)$ from $\enclaveparty$: \\
\t send $(m,r)$ to the blockchain, denote the including block $B_i$ \\
\t send a subchain from $CB$ to $B_{i+n_c}$ (inclusive) to $\enclaveparty$
}
\caption{Proof of Publication}
\label{fig:proofofpublication}
\end{figure}

\section{Proof of Main Theorem}
\label{sec:proof1}

Here we give our proof of \Cref{thm:main}, given in \Cref{sec:model}.

We prove that $\protoekiden[\lambda, \aesys, \sesys, \Sigma, \set{\ekdpartyi}_{i\in[N]}]$ UC-realizes the ideal functionality $\fekiden[\lambda, \ell, \set{\ekdpartyi}]$ with respect to a leakage function $\ell(x)$ that only reveals the length of $x$, i.e. $\ell(x)=0 ^ {|x|}$.
In the protocol, $\ell(\cdot)$ is realized with IND-CPA encryption schemes.

\begin{proof}

Let $\ekdenv$ be an environment and $\ekdadv$ be a ``dummy adversary''~\cite{ucCanetti} who simply relays messages between $\ekdenv$ and parties. To show that $\protoekiden$ UC-realizes $\fekiden$, we specify below a simulator $\simulator$ such that no environment can distinguish an interaction between $\protoekiden$ and $\ekdadv$ from an interaction with $\fekiden$ and $\simulator$, i.e. $\simulator$ satisfies
\[
\forall \ekdenv, \text{EXEC}_{\protoekiden, \ekdadv, \ekdenv} \approx \text{EXEC}_{\fekiden, \simulator, \ekdenv}.
\]

\paragraph{Construction of $\simulator$}

$\simulator$ generally proceeds as follows: if a message is sent by an honest party to $\fekiden$, $\simulator$ emulates appropriate real world ``network traffic'' for $\ekdenv$ with information obtained from $\fekiden$. If a message is sent to $\fekiden$ by a corrupted party, $\simulator$ extracts the input and interacts with the corrupted party with the help of $\fekiden$. We provide further details on the processing of specific messages.

\subparagraph{(1) Contract creation:}

\begin{itemize}
    \item If $\ekdpartyi$ is honest, $\simulator$ obtains $(\ekdpartyi, \ekdcid, \ekdprog)$ from $\fekiden$ and emulates an execution of the $\msgcreate$ call of $\protoekiden$.
    \item If $\ekdpartyi$ is corrupted, $\simulator$ extracts $\ekdprog$ from $\ekdenv$. On behalf of $\ekdpartyi$, $\simulator$ sends $(\msgcreate, \ekdprog)$ to $\fekiden$ and instructs $\fekiden$ to deliver the output.
    \item In both cases, $\simulator$ simulates the interaction between $\fstorage$ and $\funcsgx$, on behalf of the adversary or honest parties.
\end{itemize}

\subparagraph{(2) Query execution:}

\noindent \textbf{Case 1}: When an \emph{honest} party $\ekdpartyi$ is given input $(\msgquery, \ekdcid, \ekdinput, \enclaveid)$ by $\ekdenv$,
$\simulator$ works as follows: 

\begin{itemize}
    \item Upon receiving $(\ekdcid, \ekdpartyi,\ell(\ekdinput))$ from $\fekiden$, $\simulator$ queries the $\msgread$ interface of $\fekiden$ to obtain the dummy state (i.e. a random string with the same length as the real state) of $\ekdcid$, denoted $s$.
    $\simulator$ computes $c_\ekdinput=\enc(\pubkeyinput,\vec{0})$ with length $\ell(\ekdinput)$,
    and emulates a $\msgresume$ message to $\funcsgx$ with input 
    $(\msgquery, \ekdcid, c_\ekdinput, s)$ on behalf of $\ekdpartyi$.
    \item Upon receiving $\ell_{\ekdstate'}$ and $\ell(\ekdoutput)$ from $\fekiden$, $\simulator$ computes $c=\enc(\outputkey, 0^{|\ekdoutput|})$ and emulates a message $((\msginterimoutput,\hash(c_\ekdinput), \hash(s), \ell_{\ekdstate'}, \hash(c), \spk_i), \ekdatt, c)$ from $\funcsgx$ to $\ekdpartyi$.
    \item $\simulator$ proceeds by emulating the interaction between $\fstorage$ and $\funcsgx$, and a message $(\msgfinaloutput,\enc(\epk_i, 0^{|\ekdoutput|}),\ekdatt)$ from $\funcsgx$ to $\ekdpartyi$.
    \item Finally, $\simulator$ instructs $\fekiden$ by sending a $\msgok$ message.
\end{itemize}

\noindent\textbf{Case 2}: When a \emph{corrupted} party $\ekdpartyi$ is given input $(\msgquery, \ekdcid, \ekdinput, \enclaveid)$ by $\ekdenv$, $\simulator$ learns the input when  $\simulator$ works as follows:

\begin{itemize}
    \item If $\ekdpartyi$ sends $(\msgread, \ekdcid)$ to $\fstorage$, 
    $\simulator$ obtains the latest state (denoted $s$) from $\fekiden$,
    and sends $s$ to $\ekdpartyi$ on behalf of $\fstorage$. 
    \item If $\ekdpartyi$ sends a $\msgresume$ message to $\funcsgx$ with input \\
        $(\msgquery, \ekdcid, \ekdinputct, s)$, $\simulator$ emulates $\funcsgx$ as follows:
        $\simulator$ queries $\fekiden$ to check if $s$ is not the latest state,
        $\simulator$ aborts.
        $\simulator$ computes $\ekdinput'=\dec(\seckeyinput,\ekdinputct)$.
        Then $\simulator$ sends $(\msgquery, \ekdcid, \ekdinput', \enclaveid)$ to $\fekiden$ on $\ekdpartyi$'s behalf.
    \item Upon receiving $\ell_{\ekdstatect'}$ and $\ell(\ekdoutput)$ from $\fekiden$,
        $\simulator$ computes $c = \enc(\outputkey, 0^{|\ekdoutput|})$
        and sends \\ $((\msginterimoutput,\hash(\ekdinputct), \hash(s), \ell_{\ekdstatect'}, \hash(c)), \ekdatt, c)$ from $\funcsgx$ to $\ekdpartyi$.
        $\simulator$ records $c$.
    \item If $\ekdpartyi$ sends a $\msgresume$ message to $\funcsgx$ with input \\
        $(\msgclaimoutput, \ekdcid, (\ekdstatect', \ekdoutputct, \sigma, \epk_i))$, 
        $\simulator$ emulates $\funcsgx$ as follows:
        $\simulator$ first checks that $\funcsgx$ has previously sent $\ekdoutputct$ to $\ekdpartyi$ and that $(\ekdstatect', \sigma)$ has been stored by $\fstorage$.
        $\simulator$ aborts if any of the above checks fails.
        $\simulator$ obtains $\ekdoutput$ from $\fekiden$ and sends \\ $(\msgfinaloutput,\enc(\epk_i, \ekdoutput),\sigma)$ to $\ekdpartyi$.
\end{itemize}

\subparagraph{(3) Public read:}
On any call $(\msgread, \ekdcid)$ from $\ekdpartyi$,
$\simulator$ emulates a $\msgread$ message to $\fstorage$.
If $\ekdpartyi$ is corrupted, $\simulator$ sends to $\fekiden$ a $\msgread$ message
on $\ekdpartyi$'s behalf and forward the response to $\ekdadv$.

\subparagraph{(4) Corrupted enclaves:}

$\simulator$ obtains $\enclaveid$s of corrupted enclaves when $\ekdenv$ corrupts them.
In real world, $\ekdenv$ could terminate a corrupted enclave at any point, or could strategically drop some messages while letting others go through. To faithfully emulate $\ekdenv$'s ``damage'', $\simulator$ sends every messages leaving or entering a corrupted enclave to $\ekdenv$ and only delivers the message if $\ekdenv$ permits. $\simulator$ instructs $\fekiden$ to abort if the emulated execution is terminated by $\ekdenv$ prematurely. Specifically, upon receiving $(\ekdcid, \ell(\ekdstate'), \ell(\ekdoutput),\enclaveid)$ from $\fekiden$, $\simulator$ replies with $\msgok$ only if the corresponding $\msgfinaloutput$ message from $\funcsgx$ is allowed by $\ekdenv$.

\paragraph{Validity of $\simulator$}
We show that no environment can distinguish an interaction with $\adv$ and $\protoekiden$ from one with $\simulator$ and $\fekiden$ by hybrid arguments.
Consider a sequence of hybrids,
starting with the real protocol execution.
Hybrid $H_1$ lets $\simulator$ to emulate $\funcsgx$ and $\fstorage$.
$H_2$ filters out the forgery attacks against $\sgxsig$.
$H_3$ filters out the second pre-image attacks against the hash function.
$H_4$ has $\simulator$ emulate the creation phase.
$H_5$ replaces the encryption of input and output with encryption of $0$,
and replaces encryption of states with random strings with the same length.
The indispensability between adjacent hybrids are shown below.

\subparagraph{Hybrid $H_1$} proceeds as in the real world protocol, except that $\simulator$ emulates $\funcsgx$ and $\fstorage$.
Specially $\simulator$ generates a key pair $\sgxkeypair$ for $\sgxsig$ and publishes $\pksgx$.
Whenever $\ekdadv$ wants to communicate with $\funcsgx$,
$\simulator$ records $\ekdadv$'s messages and faithfully emulates $\funcsgx$'s behavior.
Similarly, $\simulator$ emulates $\fstorage$ by storing items internally.

As $\ekdadv$'s view in $H_1$ is perfectly simulated as in the real world,
$\ekdenv$ cannot distinguish between $H_1$ and the real execution.

\subparagraph{Hybrid $H_2$} proceeds as in $H_1$, except for the following modifications.
If $\adv$ invoked $\funcsgx$ with a correct message $(\msginstall, \protowrapper{\ekdprog})$,
then for all sequential $\msgresume$ calls,
$\simulator$ records a tuple $(\ekdoutput, \ekdatt)$ where $\ekdoutput$ is the output of
$\protowrapper{\ekdprog}$ and $\ekdatt$ is an attestation under $\sksgx$.
Let $\Omega$ denote the set of all such tuples. Whenever $\ekdadv$ sends an attested output $(\ekdoutput,\ekdatt)\not\in\Omega$ to $\fstorage$ or an honest party $\ekdpartyi$, $\simulator$ aborts.

The indistinguishability between $H_1$ and $H_2$ can be shown by the following reduction
to the the EU-CMA property of $\Sigma$:
In $H_1$, if $\ekdadv$ sends forged attestations to $\fstorage$ or $\ekdpartyi$,
signature verification by $\fstorage$ or an honest party $\ekdpartyi$ will fail with all but negligible probability.
If $\ekdenv$ can distinguish $H_2$ from $H_1$,
$\ekdenv$ and $\ekdadv$ can be used to win the game of signature forgery.

\subparagraph{Hybrid $H_3$} is the same as $H_2$ besides the following modifications.
If $\ekdadv$ invoked $\funcsgx$ with a correct $\msgquery$ message,
$\simulator$ records execution result $\ekdoutputct$ before outputting it.
Whenever $\ekdadv$ sends to $\funcsgx$ a $\msgclaimoutput$ message with
a input $\ekdoutputct'$ that is not previously generated by $\funcsgx$,
$\simulator$ aborts.

The indistinguishability between $H_3$ and $H_2$
can be shown by a reduction to the second pre-image resistance property of the hash function.
In $H_2$, $\ekdadv$ obtains $\mathcal{H}=\set{\hash(\ekdoutputct^i)}_i$ and
$\mathcal{O}=\set{\ekdoutputct^i}_i$ from $\funcsgx$ through $\msgquery$ calls.
If $\ekdadv$ sends a $\msgclaimoutput$ message with $\ekdoutputct\not\in \mathcal{O}$,
$\funcsgx$ aborts unless a $\hash(\ekdoutputct)\in \mathcal{H}$.
If $\ekdenv$ can distinguish $H_3$ from $H_2$,
it follows that $\ekdadv$ can break the second pre-image resistancy.

\subparagraph{Hybrid $H_4$} is the same as $H_3$ but has $\simulator$
emulate the contract creation,
i.e. honest parties will send $\msgcreate$ to $\fekiden$.
$\simulator$ emulates messages from $\funcsgx$ and $\fstorage$ as described above.
If $\ekdpartyi$ is corrupted,
$\simulator$ sends $(\msgcreate, \ekdprog)$ to $\fekiden$ as $\ekdpartyi$.

It is clear that the $\ekdadv$'s view is distributed exactly as in $H_3$,
as $\simulator$ can emulate $\funcsgx$ and $\fstorage$ perfectly.

\subparagraph{Hybrid $H_5$} is the same as $H_4$ except that 
honest parties also sends $\msgquery$ messages to $\fekiden$.
If $\ekdpartyi$ is corrupted, $\simulator$ emulates
real-world messages with the help of $\fekiden$, as described above.

In $\ekdadv$'s view, the difference between $H_5$ and $H_4$ are the following.

\begin{itemize}
\item
Any message $(\msginterimoutput,\ekdinputhash, \ekdstateprevhash, s, \ekdoutputhash, c)$ sent from
$\funcsgx$ to $\ekdpartyi$
with $s=\sesys.\enc(\seckeycont, \ekdstate')$ and $c=\sesys.\enc(\outputkey,\ekdoutput))$
in $H_4$ is replaced with $(\msginterimoutput,\ekdinputhash, \ekdstateprevhash, \ell_{\ekdstatect'}, \hash(c'), c')$
where $c'=\enc(\outputkey, 0^{|c|})$. Recall that $\ell_{\ekdstatect'}$ is a random string with length $|\ekdstatect'|$ chosen by $\fekiden$ when generating state $\ekdstatect$.

\item
If $\ekdpartyi$ is an honest party, 
any message $(\msgquery, \ekdcid, \aesys.\enc(\pubkeyinput, \ekdinput),s)$
sent to $\funcsgx$ is replaced with $(\msgquery, \ekdcid, c',s)$
where $c'=\enc(\pubkeyinput,0)$,
and
any message $(\msgfinaloutput, \aesys.\enc(\outputkey, \ekdoutput))$
sent from $\funcsgx$ to $\ekdpartyi$ is replaced with $(\msgfinaloutput,\enc(\epk_i, 0))$.
\end{itemize}

Indistinguishability between $H_5$ and $H_4$ can be directly reduced
to the IND-CPA property of $\aesys$ and $\sesys$.
Having no knowledge of the secret key, $\ekdadv$ cannot distinguish encryption of $\vec{0}$
from encryption of other messages.
Note that we don't require IND-CCA security because $\ekdadv$ do not have direct access to an decryption oracle.

It remains to observe that $H_5$ is identical to the ideal protocol.
Throughout the simulation, we maintain the following invariant:
\textbf{$\fekiden$ always has the latest state},
regardless who created the contract and who has queried the contract.
This invariant ensures that $H_5$ precisely reflects ideal execution of $\fekiden$.
\end{proof} 

\section{Ekiden Performance Extensions}
\label{sec:fullprotocol}

In this section we discuss several performance optimizations to the simple protocol.
Together, these optimizations reduce the number of round trips and storage capacity required from the blockchain,
and reduce work for compute nodes.
As we show in \Cref{sec:evaluation}, the impact is significant, up to 200\% better for write-heavy workloads.
Despite the performance improvements, all optimizations are transparent to the security interface: we use the same ideal functionality for both the simple and extended protocols. 
We present a formal protocol block defining the enhanced protocol $\protoekidenfull$ in \Cref{fig:enhancedprotocol}.
For now, we provide a high-level description of the insight and challenges involved in each application.

\paragraph{Using a write-ahead log}
In the original protocol, the entire encrypted state $\ekdstatect$ is written to the blockchain after each query.
The entire state needs to be re-encrypted because the modification side-effect should not leak information to the adversary.
However, this approach is inefficient when each $\ekdstate$ is very large yet each query modifies only a small part.
In our Token application, for example, we model a token with 500,000 different user accounts, even though each transaction only debits one account and credits one other.

Our first observation is that the use of a write-ahead log can reduce this expense. We modify the protocol so that only the ``diff'' of the state, $\Delta \ekdstatect$ is written to the blockchain.
To determine the current state, the enclave must parse the entire diff sequence, starting from the initial state, and applying each patch.
In the token application, each transaction touches a constant number of records, hence requiring $O(M+T)$ storage complexity for $T$ transactions if there are $M$ users, compared to $O(MT)$ in the simple protocol. 

The encryption of the diff $\Delta\ekdstatect$ may leak information about which query was invoked. 
The token application has constant-time queries, but in general  applications, it may be necessary to bound the size of queries and pad the ciphertext.
Finally, we note that the ideal functionality $\fekiden$ is parameterized by a leakage function $\ell$, such that the notation is in place to model the effect leakage resulting from unpadded queries.

\paragraph{Caching intermediate states at the enclave}
In the simple protocol, each round begins with reading the state ciphertext from the blockchain, and ends with writing the next state ciphertext from the blockchain. In the case that  
In our extended protocol, we optimistically use the previous state in the $\ekdcache$, if available. This results in a performance improvement when the same enclave $\enclaveid$ is used for multiple sequential queries. This is especially beneficial when the write-ahead log grows large.

Bootstrapping from genesis seems to be necessary whenever a query is sent to a new enclave 
(e.g., because the previously-used enclave host has crashed). 
In practice, we also define a policy for checkpoints by storing the entire state (not just the diff)
after every fixed number of intervals. 
We leave the formal presentation of this generalization to future work.

\paragraph{Batching transactions off-chain}
Just as the caching optimization above removes the need to read from the blockchain in each query, 
we can also coalesce the writes for multiple sequential queries into a single message to the blockchain.
This reduces both the number of network round trips, as well as the total communication cost.
When multiple queries in a batch write to the same location, only the last write needs to be stored on the blockchain.

In our protocol we do not define a policy for how many transactions must go in a batch.
Instead, we formally expose this choice to the adversary.
The choice of batching strategy has no impact on the security guarantees of our formalism. 
Each \texttt{query} invocation simply stores the inputs in a buffer, 
and the adversary can invoke the \texttt{commitBatch} method at any time to commit the entire buffer.

Batching is not a panacea. In order to maintain security, the \emph{decrypted} outputs must not 
leave the enclave unless the updated state $\Delta \ekdstatect$ is committed in the blockchain. 
Hence a user cannot receive output from a query until the entire batch is committed, and so 
only input-independent queries can appear in the same batch.

\paragraph{Coordinating the choice of compute nodes}

The Ekiden protocol leaves it up to the client to decide which compute node and enclave to query.
All of the security guarantees of $\fekiden$ hold regardless of this choice.
As a pragmatic solution, we propose to have clients defer to centralized \emph{coordinators} that perform load balancing and 
random assignment of compute nodes to tasks, based on reputations and prior experience.
If a task is not completed after some timeout, the coordinator can signal the client to repeat the query at another enclave.  
Randomization can ensure that a host cannot adaptively choose a particular target task to degrade service.
In this way Ekiden would prevent an adversary from degrading service for targeted applications. 
Following other work, incentives can be aligned by having compute miners make security deposits before they are assigned to a task.

\subsection{Extended Protocol}

An extended protocol with performance optimizations is specified in~\cref{fig:enhancedprotocol}, using the enclave program in~\cref{fig:enhancedprotocol-enclave} as a subroutine.

\begin{figure}
    \centering
    \protocol{$\protoekidenfull(\set{\ekdpartyi}_{i\in[N]})$}{
    \underline{\bf Clients $\ekdpartyi$:} \\
Initialize: $(\ssk_i,\spk_i)\sample \Sigma.\kgen(\secparam)$, $(\esk_i,\epk_i)\sample \aesys.\kgen(\secparam)$ \\[1mm]
\oninput $(\msgcreate, \ekdprog)$ from environment $\ekdenv$: \\
    \t $\ekdcid := \text{create}(\ekdprog)$ \\
    \t assert $\ekdcid$ has been stored on $\fstorage$ \\
    \t output $(\msgreceipt, \ekdcid)$ \\[1mm]
\oninput $(\msgquery, \ekdcid, \ekdinput, \enclaveid)$ from environment $\ekdenv$: \\
    \t obtains $\pubkeyinput$ from $\fstorage$ \\
    \t let $\ekdinputct := \aesys.\enc(\pubkeyinput, \ekdinput)$ \\
    \t $\sigma_{\ekdpartyi} := \sig(\ssk_i, (\ekdcid, \ekdinputct))$ \\
    \t $(\ekdstatediffct, \ekdoutputct, \sigma):=\text{query}(\ekdcid, \ekdinputct, \sigma_{\ekdpartyi})$ \\
    \t parse $\sigma$ as $(\ekdatt, \ekdinputhash, \ekdstateprevhash, \ekdoutputhash,\spk_i)$ \\
    \t assert $\sigma$ verifies \\
    \t assert $\exists n~s.t.~\ekdinputhash^n=\hash(\ekdinputct)$ \\
    \t $o:=\text{claim-output}(\ekdcid, \ekdstatediffct, \ekdoutputct, \sigma, \epk_i)$ \\
    \t \pccomment{if the previous state has been used by a parallel query} \\
    \t \pcif $o=\bot \pcthen$: jump to the beginning of this call \\
    \t parse $o$ as $(\ekdoutputct', \ekdatt)$ \\    
    \t assert $\sgxsig.\verify(\pksgx, \ekdatt, \ekdoutputct')$ \pccomment{$\pksgx := \funcsgx.\text{getpk()}$}\\
    \t output $\aesys.\dec(\esk_i, \ekdoutputct')$ \\[1mm]
\onrecv $(\msgcommitbatch, \ekdcid, \enclaveid)$ from $\ekdadv$: \\
    \t \pccomment{optimistically commit a batch without providing state} \\
    \t send $(\enclaveid, \msgresume, (\msgcommitbatch, \ekdcid, \bot))$ to $\funcsgx$ \\
    \t \pcif receive $(\msgcachemiss)$ from $\funcsgx$ \pcthen \\
    \t\t send $(\msgread, \ekdcid)$ to $\fstorage$ \\
    \t\t receive $\ekdvalue$ from \fstorage \\
    \t\t send $(\enclaveid, \msgresume, (\msgcommitbatch, \ekdcid, \ekdvalue))$ to $\funcsgx$ \\[1mm]
\onrecv $(\msgread, \ekdcid)$ from environment $\ekdenv$: \\
    \t send $(\msgread, \ekdcid)$ to $\fstorage$ \\
    \t receive $\ekdvalue$ from $\fstorage$ and $\pcreturn \ekdvalue$ \\[1mm]
\underline{\bf Compute Node Subroutines (called by $\ekdpartyi$):}\\
\oninput create($\ekdprog$):\\
    \t send $(\msginstall, \protowrapper{\ekdprog})$ to $\funcsgx$, wait for $\enclaveid$ \\
    \t send $(\enclaveid, \msgresume, (\msgcreate))$ to $\funcsgx$ \\
    \t wait for $((\ekdprog, \ekdcid, \ekdstate_0, \pubkeyinput), \ekdatt)$ from $\funcsgx$ \\
    \t send $(\msgwrite, \ekdcid, (\ekdprog, \ekdcid, \ekdstate_0, \pubkeyinput))$ to $\fstorage$ \\
    \t receive $(\msgreceipt, \ekdcid)$ from $\fstorage$ and \pcreturn \\[1mm]
\oninput query$(\ekdcid, \ekdinputct, \sigma_{\ekdpartyi})$:\\
    \t send $(\msgread, \ekdcid)$ to $\fstorage$ and wait for $\ekdstatect$ \\
    \t send $(\enclaveid, \msgresume, (\msgquery, \ekdcid, \ekdinputct, \sigma_{\ekdpartyi}, \ekdstatect))$ to $\funcsgx$ \\
    \t receive $((\ekdinputhash, \ekdstateprevhash,\ekdstatediffct, \ekdoutputhash, \spk_i), \ekdatt, \ekdoutputct)$ from \funcsgx \\
    \t let $\sigma:=(\ekdatt, \ekdinputhash, \ekdstateprevhash, \ekdoutputhash, \spk_i)$ \\
    \t \pcreturn $(\ekdstatediffct, \ekdoutputct, \sigma)$ \\[1mm]
\oninput claim-output$(\ekdcid, \ekdstatediffct, \ekdoutputct, \sigma, \epk_i)$: \\
    \t send $(\msgwrite, \ekdcid, (\ekdstatediffct, \sigma))$ to $\fstorage$ \\
    \t $\pcif$ receive $(\msgreject, \ekdcid)$ from $\fstorage$: $\pcreturn \bot$ \\
    \t send $(\enclaveid, \msgresume, (\msgclaimoutput,\ekdstatediffct, \ekdoutputct, \sigma, \epk_i))$ to $\funcsgx$ \\
    \t receive $(\msgfinaloutput,\ekdoutputct, \ekdatt)$ from \funcsgx or abort \\
    \t \pcreturn $(\ekdoutputct, \ekdatt)$ \\
    }
\caption{Enhanced \systemname Protocol. $\mathsf{diff}(\cdot,\cdot)$ is a function that takes in two states and output the difference.}
\label{fig:enhancedprotocol}
\end{figure}

\begin{figure}
\protocol{Enclave program $\protowrapper{\ekdprog}$}{
Local state: $\ekdcache := \emptyset, \ekdbatch:=\emptyset$ \\[1mm]
\oninput (\msgcreate) \\
    \t $\ekdcid := \hash(\ekdprog)$ \\
    \t $(\pubkeyinput, \seckeyinput):= \text{keyManager}(\msginputkey)$ \\
    \t $\seckeycont := \text{keyManager}(\msgstatekey)$ \\
    \t $\ekdstate_0 := \sesys.\enc(\seckeycont, \vec{0})$ \\
    \t $\ekdcache[\ekdcid] = \ekdstate_0$ \pccomment{cache state locally} \\
    \t $\pcreturn (\ekdprog, \ekdcid, \ekdstate_0, \pubkeyinput)$ \\[1mm]
\oninput $(\msgquery, \ekdcid, \ekdinputct, \sigma_{\ekdpartyi}, \ekdstatect)$ from $\ekdparty$: \\
    \t assert $\Sigma.\verify(\spk_i, \sigma_{\ekdpartyi}, (\ekdcid, \ekdinputct))$ \\
    \t add $(\ekdinputct, \spk_i)$ to $\ekdbatch[\ekdcid]$\\[1mm]
\oninput $(\msgcommitbatch, \ekdcid, \ekdinput)$: \\
    \t make a local copy of $\ekdbatch$ and parse it as $\set{(\ekdinputct^i, \spk_i)}_{i\in[N]}$ \\
    \t reset the global batch: $\ekdbatch = \emptyset$ \\
    \t \pccomment{retrieve $\pubkeyinput, \seckeyinput, \seckeycont$ from keyManager as above} \\
    \t $\ekdinput_i := \aesys.\dec(\seckeyinput, \ekdinputct^i)$ for $i\in[N]$\\
    \t \pcif $\ekdcache[\ekdcid] = \bot \land \ekdinput = \bot \pcthen:$\\
    \t\t return $(\msgcachemiss)$ \\
    \t \pcif $\ekdcache[\ekdcid] = \bot \pcthen:$ \\
    \t\t send $(\msgcheckreceipt, \ekdcid, \ekdinput)$ to $\fstorage$; wait for $\true$ or abort \\
    \t\t parse $\ekdinput$ as $\ekdstatect^0~\|~\set{\ekdstatediffct^n}_n$ \\
    \t\t reconstruct latest state and store it at $\ekdcache[\ekdcid]$ \\
    \t $\outputkey := \text{keyManager}(\msgoutputkey)$ \\
    \t let $\ekdstate[0] = \ekdcache[\ekdcid]$ \\
    \t \pcfor $i=1\dots N$:\\
    \t\t $\ekdstate[i], \ekdoutput[i] = \ekdprog(\ekdstate[i-1], \ekdinput_i, \pk_i)$ \\
    \t\t $\ekdoutputct[i] = \sesys.\enc(\outputkey, \ekdoutput[i])$ \\
    \t $\ekdcache[\ekdcid] = \ekdstate[N]$ \pccomment{cache the latest state} \\
    \t $\ekdstatediff := \mathsf{diff}(\ekdstate[N],\ekdstate[0])$ \\
    \t $\ekdinputhash := :=\hash(\ekdinputct[1])~\|~\cdots~\|~\hash(\ekdinputct[N])$ \\
    \t $\ekdstateprevhash := \hash(\ekdstate[0])$ \\
    \t $\ekdoutputhash := \hash(\ekdoutputct[1])~\|~\cdots~\|~\hash(\ekdoutputct[N])$ \\
    \t $\ekdstatediffct := \sesys.\enc(\seckeycont, \ekdstatediff)$ \\
    \t $\ekdoutputct := \ekdoutputct[1]~\|~\cdots~\|~\ekdoutputct[N]$ \\
    \t send $((\ekdinputhash,\ekdstateprevhash,\ekdstatediffct, \ekdoutputhash, \spk_i), \ekdoutputct)$ to all $\set{\ekdpartyi}_{i\in[N]}$ \\[1mm]
\oninput $(\msgclaimoutput, \ekdstatediffct, \ekdoutputct, \sigma, \epk_i)$: \\
    \t parse $\sigma$ as $(\ekdatt, \ekdinputhash, \ekdstateprevhash,\ekdoutputhash,\spk_i)$ \\
    \t parse $\ekdoutputhash$ as $\ekdoutputhash^1~\|\cdots~\|~\ekdoutputhash^n$ \\
    \t assert $\exists n~s.t.~\ekdoutputhash^n = \hash(\ekdoutputct)$ \\
    \t send $(\msgcheckreceipt, \ekdcid, (\ekdstatediffct, \sigma))$ to $\fstorage$ \\
    \t receive $\true$ from $\fstorage$ \\
    \t $\outputkey := \text{keyManager}(\msgoutputkey)$ \\
    \t $\ekdoutput := \sesys.\dec(\outputkey, \ekdoutputct)$ \\
    \t $\pcreturn (\msgfinaloutput,\aesys.\enc(\epk_i, \ekdoutput))$ \pccomment{reveal the output}
}
\caption{The enclave program used by the enhanced \systemname Protocol.}
\label{fig:enhancedprotocol-enclave}
\end{figure}

 \fi

\end{document}